# A Time-Symmetric Formulation of Quantum Measurement : Reinterpreting the Arrow of Time as Information Flow

Shin-ichi Inage

Ishinomaki Senshu University, 1 Shin-mito, Minami-sakai, Ishinomaki-shi, Miyagi-ken, 986-8580, Japan

**Abstract**

This study proposes a time-symmetric formulation of quantum measurement that restores reversibility without violating causality or thermodynamic consistency. Unlike conventional collapse-based interpretations, the measurement process is described as a bidirectional information update between a forward state $\rho_t$ and a backward effect $E_t$, governed by the completely positive generator $L$ and its adjoint $L^\dagger$. The probability $p(m \mid i, f) = \mathrm{Tr}[E\, I_m(\rho)] / \sum_k \mathrm{Tr}[E\, I_k(\rho)]$ unifies pre- and post-selected statistics under a single operator framework. This formulation preserves complete positivity, normalization, and no-signalling, and naturally satisfies Spohn's inequality, ensuring non-negative entropy production. Experimental correspondence is demonstrated for weak measurement, EPR–Bell tests, homodyne detection, and photon counting, while the classical limit converges to Kalman and RTS smoothers. Thus, temporal asymmetry arises not from physical irreversibility but from informational conditioning— the direction of knowledge update. The arrow of time is thereby redefined as the arrow of information.

**Nomenclature**

Latin letters

| | |
|---|---|
| A, B | Subsystems (e.g., Alice and Bob in bipartite systems) |
| E | Effect operator representing backward (post-selected) information |
| $E_f$ | Final effect at the terminal time t_f |
| $E_t$ | Backward-evolving effect operator under adjoint dynamics |
| H | Hamiltonian operator |
| $I_m$ | Measurement instrument corresponding to outcome m |
| $I_\Delta$ | Operation-valued measure over measurable set $\Delta$ |
| $J_Q$ | Heat flux between system and reservoir |
| $K\{m_j\}$ | Kraus operator for outcome m and component j |
| L, $L_j$ | Lindblad generator and its dissipative components |
| $L^\dagger$ | Adjoint (Heisenberg-picture) generator |
| $M_m$ | Measurement operator corresponding to outcome m |
| $\dot{Q}_t$ | Heat transfer rate at time t |
| $\Sigma(\rho_t)$ | Entropy production rate |
| $S(\rho)$ | von Neumann entropy $-\mathrm{Tr}[\rho \ln \rho]$ |
| $\dot{S}(\rho_t)$ | Time derivative of von Neumann entropy |
| T | Temperature or time-ordering operator (context-dependent) |
| Tr | Trace over the system Hilbert space |
| $U, U_t$ | Unitary time-evolution operator |
| $\dot{W}(t)$ | Power or work rate $\mathrm{Tr}[\rho_t \dot{H}(t)]$ |
| $\Delta t$ | Infinitesimal time interval |
| $\langle H \rangle_t$ | Expectation value of the Hamiltonian $\mathrm{Tr}[\rho_t H]$ |

Greek letters

| | |
|---|---|
| $\beta$ | Inverse temperature ($1/k_B T$) |
| $\beta_r$ | Inverse temperature of the r-th reservoir |
| $\gamma, \Gamma$ | Decay rate or damping constant in Lindblad form |
| $\eta$ | Measurement efficiency (e.g., in homodyne detection) |
| $\kappa$ | Coupling constant or measurement strength |
| $\lambda_i$ | Eigenvalues of a Hermitian operator (often of Hessian or density matrix) |
| $\rho$ | Density operator representing quantum state |
| $\rho, \rho_t$ | Density operator (quantum state) at time t |
| $\rho_i, \rho_f$ | Initial and final states (boundary conditions) |
| $\sigma$ | Stationary state of the Lindblad semigroup $L(\sigma)=0$ |
| $\Sigma$ | Entropy production (rate or total) |
| $\tau$ | Characteristic timescale or correlation time |
| $\psi$ | Quantum state vector (wave function) |
| $\Phi$ | Quantum channel or dynamical map (CPTP map) |
| $\Lambda$ | Completely positive (CP) pre-processing map |
| $\Omega$ | Outcome or measurable domain of an instrument |
| $\Phi_t$ | Dynamical semigroup (forward CPTP map) |
| $\Phi_{t\dagger}$ | Adjoint (Heisenberg) semigroup |
| $\Pi$ | Projection operator (used in ABL-type measurements) |
| $\Pi_m$ | Projector for measurement outcome m |

Operators and functions

| | |
|---|---|
| $[A, B]$ | Commutator $AB - BA$ |
| $\{A, B\}$ | Anti-commutator $AB + BA$ |
| $D(\rho \| \sigma)$ | Quantum relative entropy $\mathrm{Tr}[\rho(\ln\rho - \ln\sigma)]$ |
| $e^{tL}$ | Dynamical semigroup generated by L |
| ln | Natural logarithm operator acting on positive operators |
| $\mathrm{Tr}[\cdot]$ | Trace over the system Hilbert space |
| $\langle \cdot, \cdot \rangle$ | Hilbert–Schmidt inner product $\mathrm{Tr}[A^\dagger B]$ |

**Introduction**

The quest for a unified theory of time-symmetric quantum measurement and information flow has advanced intermittently for more than half a century. Originating with the time-symmetric quantum mechanics proposed by Aharonov, Bergmann, and Lebowitz (ABL) in the 1960s—known as the Two-State Vector Formalism (TSVF)—this line of research introduced an innovative perspective in which quantum measurements are conditioned simultaneously on both pre- and post-selected states. The ABL formalism provided the first explicit mathematical expression of the atemporal nature of quantum probabilities. However, the framework developed by ABL and later extended by Vaidman and Aharonov was built primarily on unitary time evolution in closed systems, and therefore could not be generalized consistently to open systems or non-unitary measurement processes involving environmental interactions. TSVF also contained interpretational ambiguities—such as treating the "bra" vector as a retro-propagating wave from the future—that, while intuitively suggestive, left unresolved issues regarding compatibility with causality and the second law of thermodynamics.

Cramer's transactional interpretation (1986) sought to account for temporal symmetry through the superposition of advanced and retarded waves in the spirit of Wheeler–Feynman absorber theory. Yet by positing a physically real "confirmation wave" arriving from the future, it conflicted with relativistic causality and lacked a clear operator formalism. These early approaches thus provided valuable intuition regarding temporal symmetry but fell short of achieving physical consistency and operational rigor.

From the 1970s onward, the operational formulation of quantum measurement advanced through the work of Davies and Ozawa, who introduced the concept of instruments to replace idealized projective measurements. Within this framework, measurement operations are represented as completely positive trace-preserving (CPTP) maps using Kraus operators, embedding measurement within the dynamics of quantum evolution itself. This reconceptualization allowed measurements to be understood not as probabilistic "outcomes" but as physical operations transforming quantum states, laying the groundwork for the modern theory of open quantum systems and continuous measurement. The dynamics of open systems were subsequently generalized by the Gorini–Kossakowski–Sudarshan–Lindblad (GKSL) equation, in which Lindblad operators generate Markovian processes that preserve complete positivity. Spohn (1978) proved that, under this formulation, quantum relative entropy decreases monotonically, thereby establishing the non-negativity of entropy production and providing a rigorous foundation for the quantum version of the second law of thermodynamics. This achievement enabled the irreversible evolution of quantum systems to be interpreted as information dissipation and entropy generation, although a direct connection to time-symmetric measurement had not yet been established.

In the 1990s, the introduction of weak measurement and weak values by Aharonov and Vaidman reignited discussions on non-invasive information extraction and the "intermediate reality" of quantum systems. Dressel and colleagues (2014) provided a comprehensive review organizing both experimental demonstrations and theoretical foundations, thereby establishing the empirical significance of time-symmetric inference. Building on this, Wiseman and Milburn (2009/2010) unified continuous monitoring, stochastic processes, and quantum feedback control, setting the stage for the Past Quantum State (PQS) formalism introduced by Gammelmark, Julsgaard, and Mølmer (2013). In the PQS framework, a forward-evolving density matrix $\rho(t)$ and a backward-evolving effect operator $E(t)$ jointly determine the conditional probability of a measurement outcome at any intermediate time, given the complete measurement record extending both forward and backward in time. This approach operationally generalized the ABL pre- and post-selection scheme, replacing the notion of a "signal from the future" with a backward information flow governed by an adjoint equation. Despite its conceptual breakthrough, the PQS theory still lacked a complete integration with Lindblad-type open-system dynamics and entropy production laws.

More recent work by Leifer and Pusey (2017) examined the relationship between operational time symmetry and retrocausality, asking whether a time-symmetric quantum theory could be constructed without invoking retrocausal assumptions. Their results suggest that temporal symmetry resides not in the reversal of physical time but in the structural symmetry of information updating, conceptually intersecting with QBism and information-theoretic quantum mechanics. Yet these approaches have so far remained at the level of probabilistic reinterpretation and have not demonstrated explicit dynamical consistency in terms of completely positive maps or adjoint operators.

Several challenges therefore remain unresolved. First, time-symmetric formalisms such as ABL and TSVF are limited to closed systems and cannot accommodate realistic non-unitary measurements. Second, a rigorous bidirectional dynamics that simultaneously satisfies causality, no-signalling, and thermodynamic irreversibility constraints has not been established. Third, a smooth theoretical connection between the quantum–classical limit—linking the Lindblad and Fokker–Planck equations, and between ABL probabilities and Bayesian smoothing probabilities—has yet to be demonstrated. Fourth, time symmetry has not been systematically formulated as operator adjointness rather than retrocausality.

The present work aims to unify these fragmented theories by reconstructing time-symmetric measurement as a dual structure of a completely positive map L and its adjoint L†. Specifically, we define the bidirectional dynamics of open quantum systems as

$$\dot{\rho}_t = L(\rho_t), \dot{E}_t = -L^\dagger(E_t),$$

so that the inner-product conservation

$$\frac{d}{dt}\text{Tr}[E_t \rho_t] = 0$$

guarantees normalization, causality, and no-signalling. Within this framework, the time-symmetric probability law corresponding to a measurement instrument $I_m$,

$$p(m \mid i, f) = \frac{\text{Tr}[E\, I_m(\rho)]}{\sum_k \text{Tr}[E\, I_k(\rho)]},$$

is derived as an operational generalization of the ABL rule. By introducing Spohn's inequality, we further demonstrate the monotonicity of entropy production, ensuring consistency with the second law of thermodynamics. In the classical limit, this bidirectional dynamics reduces smoothly to the Fokker–Planck equation and Bayesian smoothing probabilities, thereby bridging quantum probability theory, information theory, and thermodynamics.

In summary, this study redefines temporal symmetry not as a physical signal from the future or a retro-propagating particle but as a symmetry of information updating and operator adjointness within the algebra of quantum operations. It provides the first unified formalism that simultaneously satisfies complete positivity, causality, entropy monotonicity, and classical correspondence. By reinterpreting the arrow of time in quantum measurement as an informational rather than physical asymmetry, the present theory mathematically integrates and extends the major time-symmetric frameworks—ABL, TSVF, TI, PQS, and QBism—while establishing a new foundation for a general measurement theory applicable to open quantum systems.

## 2 Fundamental Structure

This section establishes the mathematical framework and notation that underlie the subsequent theorems and proofs. All constructions are formulated with minimal assumptions yet are valid for both finite-dimensional and separable Hilbert spaces. The treatment includes measurability, adjointness, composition, and limiting procedures, providing a uniform basis for discrete and continuous measurement processes.

We denote by **B(H)** the bounded operators on a Hilbert space H.

**States, Channels, and Adjoints**

The *state space* is the convex set of density operators

$$S(H) = \{\rho \in B(H) \mid \rho \geq 0,\ \text{Tr}\,\rho = 1\}. \tag{2.1}$$

The dual inner product on B(H) is $\langle X, Y \rangle := \text{Tr}[X^\dagger Y]$. A *quantum channel* is a completely positive and trace-preserving (CPTP) map $\Phi: B(H) \to B(H)$, representing forward time evolution. Every such map admits a Kraus representation

$$\Phi(\rho) = \sum_\mu N_\mu \rho N_\mu^\dagger,\ \sum_\mu N_\mu^\dagger N_\mu = I, \tag{2.2}$$

unique up to a unitary transformation among the Kraus operators (the Kraus gauge freedom).

The *adjoint* or *Heisenberg picture* map $\Phi^\dagger$ is defined by

$$\text{Tr}[X\,\Phi(Y)] = \text{Tr}[\Phi^\dagger(X)\,Y], \forall X, Y. \tag{2.3}$$

Trace preservation of $\Phi$ is equivalent to unit preservation of its adjoint, $\Phi^\dagger(I)=I$. Complete positivity is preserved under adjointness (UCP property). For any pair of times $t_1 < t_2$, we write

$$\rho(t_2) = \Phi_{t_2 \leftarrow t_1}(\rho(t_1)),\ \ X(t_1) = \Phi^\dagger_{t_2 \leftarrow t_1}(X(t_2)). \tag{2.4}$$

**Effects, POVMs, and Instruments**

An *effect* E is an operator satisfying $0 \leq E \leq I$. A collection $\{E_m\}$ with $\sum_m E_m = I$ constitutes a *POVM*.

A (discrete) *instrument* is a family $\{\,I_m\,\}$ of completely positive, trace-non-increasing maps

$$I_m(\rho) = \sum_\alpha M_{m\alpha}\,\rho\,M_{m\alpha}^\dagger,\ \sum_m I_m \text{ is TP}. \tag{2.5}$$

The associated POVM is $E_m = I_m^\dagger(I) = \sum_\alpha M_{m\alpha}^\dagger M_{m\alpha}$.

In measure-theoretic form, for a measurable space $(\Omega, \mathscr{F})$, the mapping $I: \Delta \mapsto I_\Delta$ satisfies

(i) each $I_\Delta$ is CP and trace-non-increasing;

(ii) countable additivity;

(iii) I_Ω is TP.

This defines an *operation-valued measure* (Davies–Ozawa), encompassing continuous measurements.

Gauge invariance under unitary mixing within each outcome, $M_{m\alpha} \to \sum_\beta u_{\alpha\beta}^{(m)} M_{m\beta}$, leaves $I_m$ and all probabilities invariant.

**Boundary Conditions and Backward Propagation of Effects**

A *pre-boundary* (initial condition) prepares $\rho_i$ at $t_i$:

$$\rho(t^-) = \Phi_{t^- \leftarrow t_i}(\rho_i). \tag{2.6}$$

A *post-boundary* at $t_f$ is represented by a positive effect $E_f$ whose informational influence propagates backward through the adjoint channel,

$$E(t^+) = \Phi^\dagger_{t_f \leftarrow t^+}(E_f). \tag{2.7}$$

Performing an instrument $\{I_m\}$ at an intermediate time t yields the *time-symmetric probability law*

$$p(m \mid i,f) = \frac{\text{Tr}[E(t^+) I_m(\rho(t^-))]}{\sum_k \text{Tr}[E(t^+) I_k(\rho(t^-))]}, \tag{2.8}$$

whose denominator equals $\text{Tr}[E(t^+)\rho(t^-)]$ and remains invariant under partition refinement. In Heisenberg form,

$$O_m := \sum_\alpha M^\dagger_{m\alpha} E(t^+) M_{m\alpha} \ (\geq 0), p(m \mid i,f) = \frac{\text{Tr}[O_m \rho(t^-)]}{\text{Tr}[(\sum_k O_k)\rho(t^-)]}. \tag{2.9}$$

Well-definedness requires $\text{Tr}[E(t^+)\rho(t^-)]>0$; otherwise the post-selection corresponds to a null event and conditional probability is undefined.

**Multitime Composition and Bayesian Chains**

For a temporal sequence $t_i<t_1<\cdots<t_n<t_f$ with intermediate instruments $\{I^{(j)}_{m_j}\}_{m_j}$ and free evolutions $\Phi_{j+1,j} = \Phi_{t_{j+1}\leftarrow t_j}$, the joint probability reads

$$p(i, \mathbf{m}, f) = \text{Tr}\,[E_f\, \Phi_{f,n} \circ I^{(n)}_{m_n} \circ \Phi_{n,n-1} \circ \cdots \circ I^{(1)}_{m_1} \circ \Phi_{1,i}(\rho_i)]. \tag{2.10}$$

Where $\Phi_{a,b} = \Phi_{t_a \leftarrow t_b}$.

The backward recursion of the effect operator is

$$E(t_j^+) = \Phi^\dagger_{j+1,j}\left(\sum_{m_{j+1}} I^{(j+1)\dagger}_{m_{j+1}}(E(t_{j+1}^+))\right). \tag{2.11}$$

Solving this recursion from $t_f$ backward reproduces the conditional law (2.8) for any subsequence. Coarse-graining of measurable partitions preserves consistency, $\sum_{m\in\Delta} I_m = I_\Delta$ and $\sum_{m\in\Delta} O_m$ accordingly. For spatially separated subsystems A,B, tensor-product locality uses $I^A \otimes \text{id}^B, \text{id}^A \otimes N^B$; commutativity ensures order independence (see § 5.2).

**Continuous-Time Limit and Generators**

A Markov semigroup satisfies $\Phi_{t+\Delta \leftarrow t} = \exp(\Delta L) + o(\Delta)$. For a bounded *Lindblad generator*

$$L(\rho) = -\frac{i}{\hbar}[H,\rho] + \sum_j \left(L_j \rho L_j^\dagger - \frac{1}{2}\{L_j^\dagger L_j, \rho\}\right), \tag{2.12}$$

the forward and backward equations become

$$\dot\rho_t = L(\rho_t), \dot E_t = -L^\dagger(E_t) = \frac{i}{\hbar}[H, E_t] + \sum_j \left(L_j^\dagger E_t L_j - \frac{1}{2}\{L_j^\dagger L_j, E_t\}\right), \tag{2.13}$$

which represent the continuous formulation of the *past quantum state* equations. Including continuous observation through operation-valued measures, one may expand over an infinitesimal interval dt as

$$I_{dy}(\rho) = \rho + \underbrace{L(\rho)\,dt}_{\text{free evolution}} + \underbrace{M_y(\rho)}_{\text{measurement update}} + o(dt). \tag{2.14}$$

For homodyne detection, $M_y$ contains the diffusive term $H[c]$; for counting processes, it yields a jump term (see § 4). On the backward side, the same noise realization applies via $M_y^\dagger$. The conserved inner product

$$\frac{d}{dt} \text{Tr}[E_t \rho_t] = 0 \tag{2.15}$$

follows immediately from the definition of the adjoint and expresses that "re-weighting by future information does not alter the physical flux," a key technical statement used in § 5.3.

**Commutative Limit and Classical Correspondence**

If ρ, E, and $M_{m\alpha}$ belong to a common commuting subalgebra, the trace reduces to a classical sum,

$$\text{Tr}[E\ I_m(\rho)] \rightarrow \sum_x \beta(x)\ L_m(x)\ \pi(x), \tag{2.16}$$

and Eq. (2.8) becomes identical to the product of the forward distribution π(x), likelihood $L_m(x)$, and backward message β(x) in Bayesian smoothing—thereby reproducing the classical limit discussed later.

**Technical Remarks: Well-Definedness, Boundedness, and Norms**

Normalization $\sum_m p(m \mid i, f) = 1$ follows automatically from the TP (Trace-Preserving) property of $\sum_m I_m$ is TP, $0 \leq p(m \mid i, f) \leq 1$ from $O^m \geq 0$ in (2.9). When $\text{Tr}[E(t^+)\rho(t^-)]=0$, the post-selection corresponds to a physically inaccessible event; the conditional probability is undefined, though analytic regularization $E_f \rightarrow E_f + \varepsilon\ I$ with $\varepsilon \rightarrow 0$ is admissible. Channels contract the trace norm, while their adjoints contract the operator norm, ensuring numerical stability and solvability in the continuous limit (data-processing inequality). By the *Naimark–Stinespring dilation theorem*, every instrument admits a unitary-plus-projection realization on an extended space H ⊗ K:

$$I_m(\rho) = \text{Tr}_K\ [(I \otimes \Pi_m)U(\rho \otimes |0\rangle\langle 0|)U^\dagger], \tag{2.17}$$

showing that Eq. (2.8) coincides with the ordinary Born rule on the enlarged Hilbert space (cf. § 5.1, Theorem 5.4).

**Summary of This Section**

The present framework defines the fundamental objects—states ρ, effects E, instruments I, channels Φ, and adjoints $\Phi^\dagger$—and establishes their bidirectional evolution. The probability of intermediate outcomes is given by Eq. (2.8), while multitime composition and continuous observation are coherently handled through Eqs. (2.11)–(2.14). In the commutative limit, the theory reproduces classical Bayesian smoothing; in the dilated representation, it coincides with the Born rule; and via the conserved inner product (2.15), it remains consistent with thermodynamic principles. On this foundation, § 3 develops the generalized ABL law and the past-quantum-state formalism, § 4 addresses experimental consistency, and § 5 establishes the unified compatibility of the framework with quantum mechanics, relativity, and thermodynamics.

## 3. Theorem System

We now consider a Hilbert space *H* with three ordered times: the initial time $t_i$, an intermediate time $t$, and the final time $t_f (t_i < t < t_f)$. Forward evolution is described by a completely positive and trace-preserving (CPTP) map $\Phi_{b \leftarrow a}$, and its adjoint $\Phi^\dagger_{b \leftarrow a}$ represents the Heisenberg picture. At the intermediate time, an instrument $\{I_m\}_m$ acts, where each $I_m$ is completely positive and trace-non-increasing, while $\sum_m I_m$ is trace-preserving. Given a pre-state $\rho_i$ and a post-effect $E_f \geq 0$, we introduce the shorthand

$$\rho(t^-) := \Phi_{t \leftarrow t_i}(\rho_i), E(t^+) := \Phi^\dagger_{t_f \leftarrow t}(E_f).$$

**Theorem 3.1 (Generalized ABL Law: Arbitrary Instruments and CPTP Evolution)**

**Statement.**

When an intermediate instrument $\{I_m\}$ is applied, the conditional probability of outcome $m$, given the pre-state $\rho_i$

and post-selection $E_f$, is

$$p(m \mid i, f) = \frac{\text{Tr}[E(t^+) \, I_m(\rho(t^-))]}{\sum_k \text{Tr}[E(t^+) \, I_k(\rho(t^-))]} = \frac{\text{Tr}[E(t^+) \, I_m(\rho(t^-))]}{\text{Tr}[E(t^+)\rho(t^-)]}. \quad (3.1)$$

(The second equality follows from completeness $\sum_k I_k = \text{id}$). We assume $\text{Tr}[E(t^+)\rho(t^-)] > 0$ so that the expression is well defined.

**Proof.**

The joint probability follows from the sequential Born rule:

$$p(i, m, f) = \text{Tr}\left[E_f \, \Phi_{t_f \leftarrow t}(I_m(\rho(t^-)))\right]. \quad (3.2)$$

Applying the adjoint relation $\text{Tr}[X \, \Phi(Y)] = \text{Tr}[\Phi^\dagger(X) \, Y]$ gives

$$p(i, m, f) = \text{Tr}\left[\Phi^\dagger_{t_f \leftarrow t}(E_f) I_m(\rho(t^-))\right] = \text{Tr}[E(t^+) \, I_m(\rho(t^-))]. \quad (3.3)$$

Hence $p(m \mid i, f) = p(i, m, f) / \sum_k p(i, k, f)$, yielding Eq. (3.1).

**Corollary 3.1 a (Positivity and Normalization).**

Since $I_m$ is CP and $E(t^+) \geq 0$, the numerator is non-negative, and $\sum_m p(m \mid i, f) = 1$ follows from completeness.

**Corollary 3.1 b (Kraus Gauge Invariance).**

For any unitary mixing of the Kraus decomposition

$$I_m(\rho) = \sum_\alpha M_{m\alpha} \rho M^\dagger_{m\alpha},$$

$$M_{m\alpha} \to \sum_\beta u^{(m)}_{\alpha\beta} M_{m\beta},$$

the value of $p(m \mid i, f)$ remains invariant (see Theorem 3.2).

**Special Case (Projective and Unitary Measurement).**

For a closed system with unitaries $U_1 = U(t, t_i)$, $U_2 = U(t_f, t)$, and projectors $P_m$,

$$p(m \mid i, f) = \frac{\|P_f U_2 P_m U_1 |\psi_i\rangle\|^2}{\sum_k \|P_f U_2 P_k U_1 |\psi_i\rangle\|^2}, \quad (3.4)$$

where $\rho_i = |\psi_i\rangle\langle\psi_i|$, $E_f = P_f$. This reproduces the original ABL rule.

**Theorem 3.2 (Heisenberg Representation and Effective Effects)**

**Statement.**

Define for each $m$

$$O_m := \sum_\alpha M^\dagger_{m\alpha} E(t^+) M_{m\alpha} \quad (\geq 0), \quad (3.5)$$

then

$$p(m \mid i, f) = \frac{\text{Tr}[O_m \rho(t^-)]}{\text{Tr}[(\sum_k O_k)\rho(t^-)]}, \quad \sum_k O_k = \sum_{k,\alpha} M^\dagger_{k\alpha} E(t^+) M_{k\alpha}. \quad (3.6)$$

**Proof.**

Using cyclicity of the trace,

$$\text{Tr}[E(t^+) \, I_m(\rho)] = \text{Tr}\left[\sum_\alpha M^\dagger_{m\alpha} E(t^+) M_{m\alpha} \rho\right] = \text{Tr}[O_m \rho].$$

The denominator follows analogously.

**Corollary 3.2 a (Gauge Invariance).**

Under the unitary mixing $M_{m\alpha} \to \sum_\beta u^{(m)}_{\alpha\beta} M_{m\beta}$, the sum $\sum_\alpha M^\dagger_{m\alpha} E M_{m\alpha}$ is invariant.

**Corollary 3.2 b (Convexity).**

Because both numerator and denominator depend linearly on $\rho$, $p(m \mid i, f)$ is stable under convex mixtures and projective refinements.

**Theorem 3.3 (Multitime Chain Rule and Recursive Backward Effects)**

**Statement.**

For times $t_i < t_1 < \cdots < t_n < t_f$ with instruments $\{I^{(j)}_{m_j}\}_{m_j}$ and free evolutions $\Phi_{j+1,j} := \Phi_{t_{j+1} \leftarrow t_j}$, the joint probability is

$$p(i, \mathbf{m}, f) = \text{Tr}\,[E_f\, \Phi_{f,n} \circ I^{(n)}_{m_n} \circ \Phi_{n,n-1} \circ \cdots \circ I^{(1)}_{m_1} \circ \Phi_{1,i}(\rho_i)]. \tag{3.7}$$

The backward effect satisfies the recursion

$$E(t_j^+) = \Phi^\dagger_{j+1,j}\left(\sum_{m_{j+1}} I^{(j+1)\dagger}_{m_{j+1}}(E(t_{j+1}^+))\right), E(t_n^+) = \Phi^\dagger_{f,n}(E_f), \tag{3.8}$$

and for any $j$, the conditional probability of the $j$-th outcome obeys the same structure as Eq. (3.1):

$$p(m_j \mid i, f, \text{others coarse-grained}) = \frac{\text{Tr}[E(t_j^+)\, I^{(j)}_{m_j}(\rho(t_j^-))]}{\sum_k \text{Tr}[E(t_j^+)\, I^{(j)}_k(\rho(t_j^-))]}. \tag{3.9}$$

**Proof.**

(1) Eq. (3.7) is the sequential Born expansion.

(2) Applying $\text{Tr}[X\Phi(Y)] = \text{Tr}[\Phi^\dagger(X)Y]$ stepwise from the end and inserting the coarse-graining $\sum_{m_{j+1}}$ yields the recursion (3.8).

(3) Keeping only one un-coarse-grained time $t_j$ gives Eq. (3.9). □

**Corollary 3.3 a (Coarse-Graining Stability).**

For any measurable partition, $\sum_{m \in \Delta} I_m = I_\Delta$; the denominator of (3.9) remains $\text{Tr}[E(t_j^+)\rho(t_j^-)]$, independent of partitioning.

**Corollary 3.3 b (Order Invariance for Commuting Instruments).**

Simultaneous commuting instruments at the same time commute under exchange; order invariance for spatially separated subsystems follows from § 5.2.

**Theorem 3.4 (Continuous-Time Limit: Past Quantum State Equations)**

**Statement.**

In the Markov limit $\Phi_{t+\Delta \leftarrow t} = \exp(\Delta L) + o(\Delta)$ with a bounded Lindblad generator $L$,

$$\dot\rho_t = L(\rho_t),\, \dot E_t = -L^\dagger(E_t),\, E_{t_f} = E_f. \tag{310}$$

If an instantaneous instrument $\{I_m\}$ acts at $t$, the conditional probability is

$$p(m \mid i, f) = \frac{\text{Tr}[E(t^+)\, I_m(\rho(t^-))]}{\sum_k \text{Tr}[E(t^+)\, I_k(\rho(t^-))]}, \tag{3.11}$$

the continuous-time limit of Eq. (3.1).

**Proof.**

(1) Forward evolution on $[t, t+\Delta]$: $\rho_{t+\Delta} = (I + \Delta L + o(\Delta))(\rho_t)$.

(2) Backward propagation: $E_t = (I + \Delta L^\dagger + o(\Delta))E_{t+\Delta}$, hence
$$(E_t - E_{t+\Delta})/\Delta = -L^\dagger(E_{t+\Delta}) + o(1) \Rightarrow \dot{E}_t = -L^\dagger(E_t).$$

The same holds for $\rho_t$.

(3) Compressing the instantaneous measurement into an infinitesimal interval and taking $\Delta \to 0$ yields (3.11).

(4) The conserved normalization follows from

$$\frac{d}{dt}\text{Tr}[E_t \rho_t] = \text{Tr}[-L^\dagger(E_t)\rho_t] + \text{Tr}[E_t L(\rho_t)] = 0. \tag{3.12}$$

**Corollary 3.4 a (Continuous Observation).**

Using the infinitesimal instrument $I_{dy} = \text{id} + L\, dt + M_y$, $M_y \sim \mathcal{H}[c]\, dW_t$ for homodyne detection and a jump term for counting processes, Eq. (3.11) recovers the smoothed forms of diffusive/jump stochastic-master equations (§ 4).

**Theorem 3.5 (Covariance under Pre- and Post-Processing)**

**Statement.**

Let $\Lambda(\rho) = \sum_a L_a \rho L_a^\dagger$ be a completely positive trace-preserving (CPTP) map (a pre-processing channel), and let the measurement instrument be given by Kraus operators $K_{mj}$ such that $\mathcal{I}_m(\rho) = \sum_j K_{m,j} \rho K_{m,j}^\dagger$. Then the bidirectional conditional probability defined in Eq. (2.9) is invariant under $\Lambda$, that is, $p(m \mid i, f; \Lambda(\rho)) = p(m \mid i, f; \rho)$.

**Proof. (Kraus/adjoint route without commutation).**

Define $E' = \Lambda^\dagger(E) = \sum_a L_a^\dagger E L_a$ and $\mathcal{I}'_m = \mathcal{I}_m \circ \Lambda$. Then:

$$\text{Tr}[E'\, \mathcal{I}'_m(\rho)] = \text{Tr}\Big[\Lambda^\dagger(E) \sum_{a,j} K_{m,j} L_a \rho L_a^\dagger K_{m,j}^\dagger\Big]$$

$$= \sum_{a,j} \text{Tr}[L_a^\dagger E L_a K_{m,j} \rho K_{m,j}^\dagger]$$

$$= \sum_{a,j} \text{Tr}[E L_a K_{m,j} \rho K_{m,j}^\dagger L_a^\dagger]$$

$$= \text{Tr}\Big[E \sum_j \big(\sum_a L_a K_{m,j}\big) \rho \big(\sum_{a'} L_{a'} K_{m,j}\big)^\dagger\Big]$$

$$= \text{Tr}[E\, \mathcal{I}_m(\Lambda(\rho))]. \tag{3.13}$$

Hence, both numerator and denominator of Eq. (2.9) remain identical, proving covariance. No commutation between $\mathcal{I}_m^\dagger$ and $\Lambda^\dagger$ is assumed or required.

**Implication.**

Statistical predictions are invariant under experimental "rewiring": pre- and post-channels can be absorbed into the adjoint representation without affecting measurable outcomes.

**Theorem 3.6 (Measure-Theoretic Extension to Operation-Valued Measures)**

**Statement.**

Let $(\Omega, \mathcal{F})$ be a measurable space and $I: \mathcal{F} \to \text{CP}$ an operation-valued measure with $I_\Omega$ trace-preserving.

Then for any $\Delta \in \mathcal{F}$,

$$P(\Delta \mid i, f) = \frac{\text{Tr}[E(t^+) I_\Delta(\rho(t^-))]}{\text{Tr}[E(t^+)\rho(t^-)]}, \tag{3.14}$$

defines a probability measure on $(\Omega, \mathcal{F})$: it is countably additive, normalized, and non-negative.

**Proof.**

(1) Non-negativity follows since $I_\Delta$ is CP and $E(t^+) \geq 0$.

(2) Normalization holds because $I_\Omega$ is TP, making the numerator equal to the denominator.

(3) Countable additivity follows from the strong continuity

$$I_{\sqcup_k \Delta_k} = \sum_k I_{\Delta_k} \text{ for disjoint } \Delta_k. \ \square$$

**Remark.**

The discrete case (3.1) is recovered for $\Omega = \{m\}$, $I_{\{m\}} = I_m$.

**Summary of Section 3**

- Theorem 3.1: Unified conditional-probability law (generalized ABL) valid for arbitrary instruments and CPTP evolution.
- Theorem 3.2: Heisenberg-picture form with effective effects $O_m$; clarifies Kraus-gauge invariance and normalization.
- Theorem 3.3: Multitime chain rule and recursive backward update; stable under coarse-graining and partial conditioning.
- Theorem 3.4: Continuous-time limit yields $\dot{\rho}_t = L(\rho_t)$, $\dot{E}_t = -L^\dagger(E_t)$; normalization conserved.
- Theorem 3.5: Covariance under pre-/post-processing; probabilities invariant under experimental equivalence.
- Theorem 3.6: Measure-theoretic extension to operation-valued measures, providing a unified treatment of discrete, continuous, and hybrid observations.

These results form the mathematical backbone for the consistency analyses in § 5 (quantum–relativistic–thermodynamic compatibility) and the experimental framework discussed in § 4.

## 4. Concrete Examples

**Setup.** Throughout this section we apply the unified time-symmetric law

$$p(m \mid i, f) = \frac{\text{Tr}\,[E(t^+) I_m\,(\rho(t^-))]}{\sum_k \text{Tr}\,[E(t^+) I_k\,(\rho(t^-))]}, \tag{4.1}$$

With

$$\rho(t^-) := \Phi_{t \leftarrow t_i}(\rho_i), E(t^+) := \Phi^\dagger_{t_f \leftarrow t}(E_f), \tag{4.2}$$

where $\{I_m\}$ is the instrument at the intermediate time $t$.

**(i) Single Qubit: Unified Treatment of Projective and Unsharp Z-Measurements**

**Setting.**

Take $\rho_i = |+\rangle\langle+|$ with $|+\rangle = (|0\rangle + |1\rangle)/\sqrt{2}$, choose the post-effect $E_f = |0\rangle\langle0|$, and neglect free

evolution ($\Phi = $ id). The unsharp Z-measurement is defined by the POVM elements

$$E_\pm = \frac{1}{2}(I \pm \eta \sigma_z), \eta \in [0,1]. \tag{4.3}$$

A **proper Kraus decomposition** that satisfies the completeness condition $\sum_\pm M_\pm^\dagger M_\pm = I$ is given by the Lüders-type operators $M_\pm = \sqrt{E_\pm}, \mathcal{J}_\pm(\rho) = M_\pm \rho M_\pm^\dagger$. Because $E_\pm = \frac{1}{2}(I \pm \eta \sigma_z)$ are positive operators, $\sqrt{E_\pm}$ can be written explicitly as $\sqrt{E_\pm} = \frac{1}{2}[(\sqrt{1+\eta} + \sqrt{1-\eta})I \pm (\sqrt{1+\eta} - \sqrt{1-\eta})\sigma_z]$. This construction guarantees $\sum_\pm M_\pm^\dagger M_\pm = \sum_\pm E_\pm = I$. Note that setting $M_\pm = E_\pm$ would yield $\sum_\pm M_\pm^\dagger M_\pm = \frac{1+\eta^2}{2}I \neq I$ for $\eta < 1$, violating the completeness condition. Hence $M_\pm = \sqrt{E_\pm}$ correctly defines the Kraus operators for the unsharp Z measurement and reduces to the projective case when $\eta \to 1$, this smoothly recovers the projective measurement with $E_+ = |0\rangle\langle 0|$ and $E_- = |1\rangle\langle 1|$.

**Numerator.**

Writing $\rho_i = \frac{1}{2}\begin{pmatrix} 1 & 1 \\ 1 & 1 \end{pmatrix}$, we obtain

$$\text{Tr}[E_f \mathcal{J}_+(\rho_i)] = \frac{1+\eta}{4}, \text{Tr}[E_f \mathcal{J}_-(\rho_i)] = \frac{1-\eta}{4}. \tag{4.4}$$

**Denominator.**

By completeness,

$$\sum_{m=\pm} \text{Tr}[E_f \mathcal{J}_m(\rho_i)] = \text{Tr}(E_f \rho_i) = \frac{1}{2}. \tag{4.5}$$

**Conditional probabilities.**

$$p(+|i,f) = \frac{(1+\eta)/4}{1/2} = \frac{1+\eta}{2}, p(-|i,f) = \frac{1-\eta}{2}. \tag{4.6}$$

**Limits and nonselective check.**

As $\eta \to 1$ (projective limit) $p(+|i,f) \to 1$;

as $\eta \to 0$ (ultra-weak limit) $p(+|i,f) \to 1/2$.

The nonselective marginal obeys

$$p(+|i) = \text{Tr}[\mathcal{J}_+(\rho_i)] = \frac{1}{2} \text{ (independent of } \eta), \tag{4.7}$$

consistent with no-signalling.

**(ii) Weak Measurement: First-Order Derivation of the Weak Value** $A_w = \langle f | A | i \rangle / \langle f | i \rangle$

**Setting.** Couple system $H_S$ and meter $H_M$ (pointer variables $q, p$ with $[q,p] = i$; we take $\hbar = 1$). The meter is initialized in a real Gaussian $|G\rangle$ with $\langle q \rangle_G = \langle p \rangle_G = 0$ and variance $\text{Var}(q) = \sigma_q^2$. An impulsive coupling at $t$ implements

$$U = \exp(-ig A \otimes p) \simeq I - ig A \otimes p - \frac{1}{2}g^2 A^2 \otimes p^2 + \cdots, g \ll 1. \tag{4.8}$$

**Post-selection and expansion.** Conditioning on $|f\rangle$ at the end,

$$\langle f | U | i \rangle | G \rangle \simeq \langle f | i \rangle (I - ig A_w p) | G \rangle, A_w := \frac{\langle f|A|i\rangle}{\langle f|i\rangle}. \tag{4.9}$$

**Pointer shifts (first order).** With $\langle [q,p] \rangle_G = i$ and $\langle \{q,p\} \rangle_G = 0$,

$$\Delta q := \langle q \rangle_f = g \, \mathrm{Re}\, A_w + O(g^2), \Delta p := \langle p \rangle_f = 2g \, \mathrm{Var}(p) \, \mathrm{Im}\, A_w + O(g^2). \tag{4.10}$$

For a real Gaussian, $\mathrm{Var}(p) = 1/(4\sigma_q^2)$, hence

$$\Delta p = \frac{g}{2\sigma_q^2} \, \mathrm{Im}\, A_w + O(g^2). \tag{4.11}$$

**Equivalence with the unified law.** Expanding the induced system instrument $I(\rho) = \mathrm{Tr}_M[U(\rho \otimes |G\rangle\langle G|)U^\dagger]$ to first order and inserting into (4.1) reproduces (4.10)–(4.11).

### (iii) Continuous Homodyne Monitoring: SME, Adjoint SDE, and Past Quantum State

**Setting.** A single damped mode with rate $\kappa$ and measurement operator $c$ is monitored by homodyne detection (local-oscillator phase chooses $X_c := c + c^\dagger$). Detection efficiency $\eta \in [0,1]$. The measured current is

$$dY_t = 2\sqrt{\eta\kappa} \, \langle X_c \rangle_t \, dt + dW_t, \tag{4.12}$$

with $dW_t$ a Wiener increment.

**Forward (SME).**

$$d\rho_t = \mathcal{L}(\rho_t) \, dt + \sqrt{\eta\kappa} \, \mathcal{H}[c](\rho_t) \, dW_t, \mathcal{L}(\rho) = -\frac{i}{\hbar}[H, \rho] + \kappa \left( c\rho c^\dagger - \frac{1}{2}\{c^\dagger c, \rho\} \right),$$

$$\mathcal{H}[c](\rho) = c\rho + \rho c^\dagger - \mathrm{Tr}\,[(c + c^\dagger)\rho]\rho. \tag{4.13}$$

**Backward (adjoint SDE for the effect).**

$$dE_t = -\mathcal{L}^\dagger(E_t) \, dt + \sqrt{\eta\kappa} \, \mathcal{H}^\dagger[c](E_t) \, dW_t,$$

$$\mathcal{L}^\dagger(E) = \frac{i}{\hbar}[H, E] + \kappa \left( c^\dagger E c - \frac{1}{2}\{c^\dagger c, E\} \right), \mathcal{H}^\dagger[c](E) = c^\dagger E + E c - \mathrm{Tr}\,[E(c + c^\dagger)]E,$$

$$\tag{4.14}$$

with terminal condition $E_{t_f} = E_f$.

**Probability of an inserted event.** If a discrete event (e.g. a projector $P$) is hypothetically inserted at time $t$, the unified law gives

$$p(P \text{ at } t \mid \text{full record}) = \frac{\mathrm{Tr}\,[E(t^+) \, P \, \rho(t^-) \, P]}{\mathrm{Tr}\,[E(t^+) \, \rho(t^-)]}, \tag{4.15}$$

and normalization follows by completeness. Thus the *smoothing* formula of continuous measurement coincides exactly with the time-symmetric law applied to the PQS pair $(\rho_t, E_t)$.

### (iv) EPR Correlations: Recovering Bell Violations and Manifest No-Signalling

**Setting.** Prepare $|\Phi^+\rangle = (|00\rangle + |11\rangle)/\sqrt{2}$. Alice (A) and Bob (B) measure $\pm 1$ along Bloch directions $\mathbf{a}$ and $\mathbf{b}$:

$$P_{\mathbf{a}}^\pm = \frac{I \pm \mathbf{a} \cdot \boldsymbol{\sigma}}{2}, P_{\mathbf{b}}^\pm = \frac{I \pm \mathbf{b} \cdot \boldsymbol{\sigma}}{2}.$$

**Joint probabilities (Born rule).** For $\rho_{AB} = |\Phi^+\rangle\langle\Phi^+|$,

$$p(\pm, \pm) = \mathrm{Tr}\,[(P_{\mathbf{a}}^\pm \otimes P_{\mathbf{b}}^\pm)\rho_{AB}] = \frac{1 \pm \mathbf{a} \cdot \mathbf{b}}{4}, p(\pm, \mp) = \frac{1 \mp \mathbf{a} \cdot \mathbf{b}}{4}. \tag{4.16}$$

Hence

$$E(\mathbf{a}, \mathbf{b}) := \sum_{s,t=\pm} st \, p(s, t) = \mathbf{a} \cdot \mathbf{b}, \tag{4.17}$$

and for optimal CHSH settings one obtains

$$S = 2\sqrt{2}. \quad (4.18)$$

**No-signalling (nonselective marginals).**

$$p_B(\pm) = \sum_{s=\pm} \text{Tr}\left[(P_\mathbf{a}^s \otimes P_\mathbf{b}^\pm)\rho_{AB}\right] = \text{Tr}\left[(I \otimes P_\mathbf{b}^\pm)\rho_{AB}\right] = \frac{1}{2}, \quad (4.19)$$

independent of Alice's choice **a** or even whether she measures. This is the concrete instance of completeness $\sum_s P_\mathbf{a}^s = I$.

**Effect of post-selection (selective).**

If Alice conditions on $s = +$,

$$p_B(\pm \mid s = +) = \frac{\text{Tr}\left[(P_\mathbf{a}^+ \otimes P_\mathbf{b}^\pm)\rho_{AB}\right]}{\sum_t \text{Tr}\left[(P_\mathbf{a}^+ \otimes P_\mathbf{b}^t)\rho_{AB}\right]} = \frac{1 \pm \mathbf{a} \cdot \mathbf{b}}{2}, \quad (4.20)$$

i.e., conditioning reshapes Bob's *subsample*, but the *nonselective* statistics (4.19) remain $\frac{1}{2}$; no signal can be sent.

**Instrument–effect rewrite.** More generally, inserting any post-effect $E_f^A$ on A and summing over A's outcomes,

$$\sum_s \text{Tr}\left[(E_f^A \otimes P_\mathbf{b}^\pm)(I_s^A \otimes \text{id}_B)(\rho_{AB})\right] = \text{Tr}\left[(\tilde{E}_f^A \otimes P_\mathbf{b}^\pm)\rho_{AB}\right], \tilde{E}_f^A := \sum_s I_s^{A\dagger}(E_f^A),$$

$$(4.21)$$

and completeness implies $\text{Tr}_A[\tilde{E}_f^A \rho_{AB}] = \text{Tr}_A[\rho_{AB}]$, so Bob's marginal remains unchanged.

**Summary of Section 4**

(i) For a single qubit, the parameter $\eta$ continuously interpolates between weak and projective regimes, and post-selection reshapes the conditional distribution exactly as (4.6) predicts while preserving nonselective marginals (4.7).

(ii) In the weak-measurement limit, the weak value $A_w$ governs the first-order pointer shifts (4.10)–(4.11), in full agreement with the unified law.

(iii) Under continuous homodyne monitoring, the *forward SME + adjoint SDE* pair (4.13)–(4.14) yields, via (4.15), a one-line smoothing formula identical to the time-symmetric rule.

(iv) For EPR states, the framework reproduces maximal CHSH violation (4.18) while manifesting no-signalling through invariant nonselective marginals (4.19); selective conditioning (4.20) changes only the post-selected subsequence.

## 5. Consistency of the Bidirectional Framework

### 5.1 Consistency with Quantum Mechanics

This section demonstrates that the proposed framework is *strictly consistent* with standard quantum mechanics, including the Born rule, unitary time evolution, POVM/instrument formalism, and Lüders update. Let the system's Hilbert space be $\mathcal{H}$, and the density operator satisfy $\rho \geq 0$, $\text{Tr}\rho = 1$. Time evolution is represented by a completely positive and trace-preserving (CPTP) map $\Phi_{t_2 \leftarrow t_1}: \rho \mapsto U(t_2, t_1)\rho U^\dagger(t_2, t_1)$, for a closed system with $U(t_2, t_1) = e^{-iH(t_2 - t_1)/\hbar}$.

An instrument $\{I_m\}_m$ is given by $I_m(\rho) = \sum_\alpha M_{m\alpha} \rho M_{m\alpha}^\dagger$, $\sum_m I_m$ is TP.

The final effect $E_f \geq 0$ evolves backward by the adjoint channel as $E(t) = \Phi_{t_f \leftarrow t}^\dagger(E_f)$. Then, the unified

conditional probability law reads:

$$p(m \mid i, f) = \frac{\text{Tr}[E(t)I_m(\rho(t))]}{\sum_k \text{Tr}[E(t)I_k(\rho(t))]}, \rho(t) = \Phi_{t \leftarrow t_i}(\rho_i). \tag{5.1}$$

**Proposition 5.1 (Born rule recovery)**

**Claim.**

If the post-selection is removed, i.e. $E_f = I$ so that $E(t) = I$, Eq. (5.1) exactly reduces to the standard Born rule $p(m \mid i) = \text{Tr}[I_m(\rho(t))] = \text{Tr}[E_m \rho(t)], E_m := \sum_\alpha M_{m\alpha}^\dagger M_{m\alpha}$, where $\{E_m\}$ form a POVM: $\sum_m E_m = I, E_m \geq 0$.

**Proof.**

Substituting $E(t) = I$ into (5.1): $p(m \mid i) = \frac{\text{Tr}[I \, I_m(\rho(t))]}{\text{Tr}[(\sum_k I_k)(\rho(t))]} = \text{Tr}[I_m(\rho(t))]$, because $\sum_k I_k$ is trace-preserving.

Using cyclicity of trace:

$$\text{Tr}[I_m(\rho)] = \text{Tr}\left[\sum_\alpha M_{m\alpha} \rho M_{m\alpha}^\dagger\right] = \text{Tr}\left[\sum_\alpha M_{m\alpha}^\dagger M_{m\alpha} \rho\right] = \text{Tr}[E_m \rho].$$

Hence $E_m \geq 0$ and $\sum_m E_m = I$.

**Proposition 5.2 (Sequential measurement consistency)**

**Claim.**

Let $\rho_i$ be prepared at $t_i$, propagated forward by $\Phi_{t \leftarrow t_i}$ and $\Phi_{t_f \leftarrow t}$, with intermediate instrument $\{I_m\}$ at $t$ and final effect $E_f$. The joint probability reads:

$$p(i, m, f) = \text{Tr}\left[E_f \Phi_{t_f \leftarrow t}(I_m(\Phi_{t \leftarrow t_i}(\rho_i)))\right], \tag{5.2}$$

and (5.1) follows as $p(m \mid i, f) = p(i, m, f)/\sum_k p(i, k, f)$. For a closed system with projectors $M_{m\alpha} = P_m$,

$$p(i, m, f) = \| P_f U_2 P_m U_1 \mid \psi_i \rangle \|^2, \tag{5.3}$$

with $U_1 = U(t, t_i)$, $U_2 = U(t_f, t)$, and $\rho_i = \mid \psi_i \rangle \langle \psi_i \mid$.

**Proof.**

By successive Born applications: $p(i, m, f) = \text{Tr}[E_f \Phi_{t_f \leftarrow t}(I_m(\Phi_{t \leftarrow t_i}(\rho_i)))]$. Using $\text{Tr}[X\Phi(Y)] = \text{Tr}[\Phi^\dagger(X)Y]$ gives $p(i, m, f) = \text{Tr}[\Phi_{t_f \leftarrow t}^\dagger(E_f)I_m(\Phi_{t \leftarrow t_i}(\rho_i))] = \text{Tr}[E(t)I_m(\rho(t))]$. Normalization yields (5.1). For projective $P_m$: $p(i, m, f) = \langle \psi_i \mid U_1^\dagger P_m U_2^\dagger P_f U_2 P_m U_1 \mid \psi_i \rangle = \| P_f U_2 P_m U_1 \mid \psi_i \rangle \|^2$.

**Proposition 5.3 (Lüders update recovery)**

**Claim.**

For projective measurement, the unconditional and conditional post-measurement states are

$$\rho \mapsto \sum_m P_m \rho P_m, \rho_m = \frac{P_m \rho P_m}{\text{Tr}[P_m \rho]}, \tag{5.4}$$

consistent with (5.1).

**Proof.**

The non-selective probability is $\mathrm{Tr}[P_m \rho]$. For selective update, $I_m(\rho) = P_m \rho P_m$ and $\rho_m = I_m(\rho)/\mathrm{Tr}[I_m(\rho)]$, giving (5.4). □

**Theorem 5.4 (Naimark extension equivalence)**

**Claim.**

Any instrument $\{I_m\}$ can be realized on an enlarged space $\mathcal{H} \otimes \mathcal{K}$ with ancilla $|0\rangle \in \mathcal{K}$, unitary $U$, and projectors $\{\Pi_m\}$:

$$I_m(\rho) = \mathrm{Tr}_\mathcal{K}[(I \otimes \Pi_m) U(\rho \otimes |0\rangle\langle 0|) U^\dagger]. \tag{5.5}$$

Then (5.1) becomes equivalent to the ordinary Born rule in the extended space. Defining $\widetilde{E_f} = E_f \otimes I_\mathcal{K}$ and $\widetilde{\rho_i} = \rho_i \otimes |0\rangle\langle 0|$,

$$\mathrm{Tr}[E(t) I_m(\rho(t))] = \mathrm{Tr}[\widetilde{E_f}\, \widetilde{\Phi}_{t_f \leftarrow t}((I \otimes \Pi_m)\widetilde{\Phi}_{t \leftarrow t_i}(\widetilde{\rho_i}))]. \tag{5.6}$$

**Proof.**

From Stinespring representation (5.5). Using cyclicity and $\mathrm{Tr}_\mathcal{H}[A\, \mathrm{Tr}_\mathcal{K}(B)] = \mathrm{Tr}_{\mathcal{H} \otimes \mathcal{K}}[(A \otimes I)B]$, Eq. (5.6) follows. Thus, (5.1) corresponds exactly to the Born rule in the extended space. □

**Proposition 5.5 (Heisenberg representation)**

Define effective effects

$$O_m := \sum_\alpha M_{m\alpha}^\dagger E(t) M_{m\alpha} \geq 0. \tag{5.7}$$

Then

$$p(m \mid i, f) = \frac{\mathrm{Tr}[O_m \rho(t)]}{\mathrm{Tr}[(\sum_k O_k) \rho(t)]}. \tag{5.8}$$

**Proof.**

$$\mathrm{Tr}[E(t) I_m(\rho)] = \mathrm{Tr}[E(t) \sum_\alpha M_{m\alpha} \rho M_{m\alpha}^\dagger] = \mathrm{Tr}[\sum_\alpha M_{m\alpha}^\dagger E(t) M_{m\alpha} \rho] = \mathrm{Tr}[O_m \rho].$$

Hence (5.8) holds.

Since $O_m \geq 0$, probabilities are well-defined and invariant under local Kraus unitaries $M_{m\alpha} \to \sum_\beta u_{\alpha\beta}^{(m)} M_{m\beta}$.

**Proposition 5.6 (Schrödinger–Heisenberg equivalence)**

For closed-system unitaries $U_1 = U(t, t_i)$, $U_2 = U(t_f, t)$:

$$\mathrm{Tr}[E(t) I_m(\rho(t))] = \mathrm{Tr}[E_f\, U_2\, I_m(U_1 \rho_i U_1^\dagger) U_2^\dagger] = \mathrm{Tr}[O_m^{(H)} \rho_i], \tag{5.9}$$

where

$$O_m^{(H)} := U_1^\dagger \left(\sum_\alpha \widetilde{M}_{m\alpha}^\dagger E_f \widetilde{M}_{m\alpha}\right) U_1, \quad \widetilde{M}_{m\alpha} := U_2 M_{m\alpha} U_2^\dagger.$$

Thus, the forward and backward representations are fully equivalent.

**Proposition 5.7 (Normalization and boundedness)**

For all $\rho(t) \geq 0$, $E(t) \geq 0$ and any instrument $\{I_m\}$, Eq. (5.1) satisfies

(i) $0 \leq p(m \mid i, f) \leq 1$,
(ii) $\sum_m p(m \mid i, f) = 1$.

**Proof.**

From (5.7), $O_m \geq 0$ and $\sum_m O_m \geq 0$.

Both numerator and denominator are non-negative, with denominator $> 0$. Hence $p(m \mid i, f) \geq 0$ and $\sum_m p(m \mid i, f) = \frac{\sum_m \text{Tr}[O_m \rho]}{\text{Tr}[(\sum_k O_k)\rho]} = 1$, and $p(m \mid i, f) \leq 1$ follows from $\text{Tr}[O_m \rho] \leq \text{Tr}[(\sum_k O_k)\rho]$.

**Proposition 5.8 (Closure under composition)**

The composed channel $\Phi_2 \circ I_m \circ \Phi_1$ is CP and trace-nonincreasing; $\sum_m (\Phi_2 \circ I_m \circ \Phi_1)$ is TP. Hence the formalism is closed under arbitrary temporal concatenations and measurements—i.e., internally consistent within the category of quantum operations.

**Proof.**

Composition of CP maps is CP. Since $I_m$ is trace-nonincreasing and $\Phi_{1,2}$ are TP, $\text{Tr}[(\Phi_2 \circ I_m \circ \Phi_1)(\rho)] \leq \text{Tr}[\rho]$. The completeness $\sum_m I_m$ being TP ensures total trace preservation.

**Summary of §5.1**

Equation (5.1) simultaneously reproduces:
- the Born rule (Prop. 5.1),
- Sequential measurement law (Prop. 5.2),
- Lüders projection update (Prop. 5.3),
- Naimark extension theorem (Thm. 5.4),
- Heisenberg backward effect (Props. 5.5–5.6),
- Normalization and boundedness (Prop. 5.7), and
- Closure in the CP-map category (Prop. 5.8).

Therefore, Eq. (5.1) constitutes an *intrinsic representation of standard quantum mechanics*—complete, self-consistent, and free from any extraneous assumptions.

**5.2 Consistency with Relativistic Causality**

We now prove that the bidirectional framework—forward state evolution $\Phi$, backward effect evolution $\Phi^\dagger$, and intermediate instruments $\{I\}$—is fully consistent with relativistic causality (no superluminal signalling) and Lorentz covariance. We work with two spacelike-separated subsystems $A$ and $B$ on $\mathcal{H}_A \otimes \mathcal{H}_B$. *Locality* means operations on $A$ belong to $B(\mathcal{H}_A) \otimes I_B$ and those on $B$ to $I_A \otimes B(\mathcal{H}_B)$.

**Theorem 5.2.1 (No-signalling in Nonselective Statistics)**

**Claim.**

For any initial state $\rho_{AB}$, any local instrument $\{I_a^A\}_a$ on $A$ with $\sum_a I_a^A$ trace-preserving, and any local instrument $\{N_b^B\}_b$ on $B$, the *nonselective* marginal on $B$ is

$$p_B(b) = \sum_a \text{Tr} \, [(I_a^A \otimes N_b^B)(\rho_{AB})] = \text{Tr}[(\text{id}_A \otimes N_b^B)(\rho_{AB})], \qquad (5.10)$$

hence independent of $A$'s choice of operation, measurement basis, or post-effect. Superluminal signalling $A \to B$ is therefore impossible.

**Proof.**

Completeness gives $\sum_a I_a^A = \text{id}_A$, hence $p_B(b) = \text{Tr}\,[(\sum_a I_a^A \otimes N_b^B)\rho_{AB}] = \text{Tr}\,[(\text{id}_A \otimes N_b^B)\rho_{AB}]$. If a post-effect $E_f^A$ is introduced on $A$ but *not* conditioned upon, then

$$\sum_a \text{Tr}\,[(E_f^A \otimes I_B)(I_a^A \otimes N_b^B)(\rho_{AB})] = \text{Tr}\,[(\tilde{E}_f^A \otimes I_B)(\text{id}_A \otimes N_b^B)(\rho_{AB})], \qquad (5.11)$$

with $\tilde{E}_f^A := \sum_a I_a^{A\dagger}(E_f^A)$. Local completeness implies $\text{Tr}_A[\tilde{E}_f^A X_A] = \text{Tr}_A[X_A]$ for all $X_A$, so the right-hand side equals $\text{Tr}[(\text{id}_A \otimes N_b^B)\rho_{AB}]$.

**Corollary.** The conclusion remains valid when $A$'s actions are time-symmetric, i.e. including backward effects $E_t^A = \Phi_A^\dagger(E_f^A)$, since $\Phi_A^\dagger$ is a *local* adjoint map and the computation remains confined to $A$'s algebra.

**Theorem 5.2.2 (Selective Statistics Do Not Enable Signalling)**

**Claim.**

Conditioning on an event $C$ on $A$ (a specific outcome or successful post-selection) may alter the *conditional* marginal $p_B(b \mid C)$. However, Bob's *nonselective* marginal satisfies the law of total probability

$$p_B(b) = p_B(b \mid C)\, p(C) + p_B(b \mid \bar{C})\, p(\bar{C}), \qquad (5.12)$$

which equals the value in (5.10). Thus $B$ cannot infer the occurrence of $C$ without classical communication.

**Proof.**

Define the effect associated with event $C$ by a CP-adjoint action $\Pi_C^A$. Then

$$p_B(b \mid C) = \frac{\text{Tr}[(\Pi_C^A \otimes N_b^B)\rho_{AB}]}{\text{Tr}[(\Pi_C^A \otimes I)\rho_{AB}]}, \qquad (5.13)$$

and summing over $C, \bar{C}$ reproduces (5.12). Using the instrument-level decomposition,

$$p_B(b) = \sum_{a \in C} \text{Tr}\,[(I_a^A \otimes N_b^B)\rho] + \sum_{a \notin C} \text{Tr}\,[(I_a^A \otimes N_b^B)\rho] = \sum_a \text{Tr}\,[(I_a^A \otimes N_b^B)\rho],$$

which equals (5.10).

**Remark (Operational meaning).** Selective post-processing reshapes *conditioned* subensembles, but announcing which subensemble requires classical communication (LOCC), bounded by light speed; hence causality is preserved.

**Theorem 5.2.3 (Order Invariance under Spacelike Separation)**

**Claim.**

For spacelike-separated local operations, the joint probability is independent of their order:

$$p(a,b) = \text{Tr}\,[(I_a^A \otimes N_b^B)\rho_{AB}] = \text{Tr}\,[(N_b^B \otimes I_a^A)\rho_{AB}]. \qquad (5.14)$$

**Proof.**

With Kraus forms $I_a^A(\cdot) = \sum_\alpha M_{a\alpha}^A (\cdot) M_{a\alpha}^{A\dagger}$ and $N_b^B(\cdot) = \sum_\beta N_{b\beta}^B (\cdot) N_{b\beta}^{B\dagger}$,

$$(I_a^A \otimes N_b^B): \rho \mapsto \sum_{\alpha,\beta} (M_{a\alpha}^A \otimes N_{b\beta}^B)\, \rho\, (M_{a\alpha}^A \otimes N_{b\beta}^B)^\dagger.$$

Since the factors act on different tensor components, $(M^A_{a\alpha} \otimes I)(I \otimes N^B_{b\beta}) = (I \otimes N^B_{b\beta})(M^A_{a\alpha} \otimes I)$. Thus the operator-sum and the resulting trace are order-independent. The same holds when backward effects $E^{(A,B)}_t = \Phi^\dagger_{(A,B)}(E^{(A,B)}_f)$ are included, as these are local adjoints. □

**Theorem 5.2.4 (Lorentz Covariance of Probabilities)**
**Claim.**

Let $U(\Lambda)$ be a unitary representation of a Lorentz transformation $\Lambda$ on the global Hilbert space, and suppose local operations transform covariantly as

$$\rho \mapsto \rho' = U\rho U^\dagger, M_{a\alpha} \mapsto M'_{a\alpha} = UM_{a\alpha}U^\dagger, E \mapsto E' = U^\dagger E U. \tag{5.15}$$

Then the basic numerator of the unified law is Lorentz invariant:

$$\mathrm{Tr}[E'\, I'_a(\rho')] = \mathrm{Tr}[E\, I_a(\rho)]. \tag{5.16}$$

**Proof.**
Using $I'_a(\rho') = \sum_\alpha M'_{a\alpha} \rho' M'^\dagger_{a\alpha}$ and cyclicity,

$$\mathrm{Tr}\left[U^\dagger E U \sum_\alpha (UM_{a\alpha}U^\dagger)(U\rho U^\dagger)(UM^\dagger_{a\alpha}U^\dagger)\right] = \mathrm{Tr}\left[E \sum_\alpha M_{a\alpha}\rho M^\dagger_{a\alpha}\right].$$

The denominator transforms identically; hence $p(m \mid i, f)$ is a scalar, independent of the inertial frame.
**Remark.** The same conclusion follows in the Tomonaga–Schwinger formalism, where evolution proceeds between spacelike hypersurfaces; probabilities are frame-independent scalars.

**Theorem 5.2.5 (Microcausality and Local CP Actions in QFT)**
**Claim.**

Let the local von Neumann algebras $\mathcal{A}(O_A)$ and $\mathcal{A}(O_B)$ be associated with spacelike-separated regions and satisfy microcausality, $[X_A, Y_B] = 0$ for all $X_A \in \mathcal{A}(O_A)$, $Y_B \in \mathcal{A}(O_B)$. If the Kraus generators of local CP maps $I^A_a, N^B_b$ lie in the respective local algebras, then the local CP actions commute and Theorems 5.2.1–5.2.3 extend verbatim.

**Proof.**

Taking $M^A_{a\alpha} \in \mathcal{A}(O_A)$ and $N^B_{b\beta} \in \mathcal{A}(O_B)$, microcausality yields $[M^A_{a\alpha}, N^B_{b\beta}] = 0$. Hence the composed actions commute, and the joint/marginal derivations are identical to the finite-dimensional case. Backward effects $E_t = \Phi^\dagger(E_f)$ remain inside the local adjoint image and preserve commutativity.

**Proposition 5.2.6 (Closure under Time Slicing and Concatenation)**
**Claim.**

Partition any finite interval into sub-intervals; in each slice, insert arbitrary *local* CPTP evolutions and local instruments. The assertions of Theorems 5.2.1–5.2.4 persist under sequential composition.

**Proof.**

Tensor products of local CPTP maps are CPTP; direct sums of local instruments are TP. The nonselective no-signalling calculation (5.10) holds slice-by-slice and is invariant under concatenation. □

**Proposition 5.2.7 (Formal Impossibility of "Postselection Signalling")**

**Claim.**

Even if Alice varies a protocol to change the *success probability* of her postselection, Bob's *nonselective* marginal remains unchanged; postselection alone cannot carry a signal.

**Proof.**

Let $\Pi_C^A$ denote the success effect on $A$, and $\sum_a I_a^A = \text{id}_A$. Then

$$p_B(b) = \sum_a \text{Tr}\left[(I_a^A \otimes N_b^B)\rho\right] = \text{Tr}[(\text{id}_A \otimes N_b^B)\rho], \qquad (5.17)$$

independent of the choice of $\Pi_C^A$. Distinguishing different protocols at $B$ requires classical information from $A$, hence no superluminal signalling. □

**Summary of §5.2**

- Nonselective invariance: Local completeness $\sum_a I_a^A = \text{id}$ guarantees that the remote nonselective marginal is unchanged (Thm. 5.2.1).
- Selectivity ≠ signalling: Conditional subensembles may differ, but average back to the same nonselective marginal (Thm. 5.2.2).
- Order invariance: Spacelike-separated local operations commute operationally; joint statistics are order-independent (Thm. 5.2.3).
- Lorentz covariance: Probabilities are frame-independent scalars under $U(\Lambda)$ (Thm. 5.2.4).
- Field-theoretic extension: Under microcausality, local CP actions commute and no-signalling extends to QFT (Thm. 5.2.5).
- Concatenation closure: The properties persist under arbitrary time slicing and composition (Prop. 5.2.6).
- No "postselection signalling": Without classical communication, postselection cannot transmit information (Prop. 5.2.7).

Hence, the bidirectional framework—*including backward effects from future post-selections*—is rigorously consistent with relativistic causality and Lorentz covariance: it neither enables superluminal communication nor introduces order dependence for spacelike-separated operations.

**5.3 Consistency with Thermodynamics and Entropy Production**

We show that the bidirectional framework—forward state equation $\dot{\rho}_t = L(\rho_t)$, backward effect equation $\dot{E}_t = -L^\dagger(E_t)$, and intermediate instruments $\{I_m\}$—is fully consistent with the second law of quantum thermodynamics. In what follows, we explicitly invoke Spohn's inequality [Spohn, *J.* 1978] to guarantee thermodynamic consistency. For any completely positive dynamical semigroup $e^{t\mathcal{L}}$ with stationary state $\rho_{ss}$ satisfying $\mathcal{L}(\rho_{ss}) = 0$, the entropy production rate $\sigma(t) = -\frac{d}{dt}S(\rho_t \parallel \rho_{ss}) = -\text{Tr}\left[(\mathcal{L}\rho_t)(\ln \rho_t - \ln \rho_{ss})\right] \geq 0$ ensures monotonic contraction of the quantum relative entropy and thus non-negative entropy generation.

**Theorem 5.3 (Spohn inequality and entropy production).**

Let $\mathcal{L}$ be a Lindblad generator with stationary state $\sigma$ satisfying $\mathcal{L}^\dagger(\sigma) = 0$. Then, for the dynamical semigroup $\Phi_t = e^{t\mathcal{L}}$ and any state $\rho$, $\frac{d}{dt}D(\Phi_t(\rho) \parallel \sigma) = \text{Tr}\left[\mathcal{L}(\Phi_t(\rho))(\log \Phi_t(\rho) - \log \sigma)\right] \leq 0$, where $D(\rho \parallel \sigma) =$

$\mathrm{Tr}[\rho(\log \rho - \log \sigma)]$ is the quantum relative entropy. Hence, the entropy production rate $\dot{\Sigma}(t) \equiv -\frac{d}{dt}D(\rho_t \parallel \sigma) \geq 0$, guaranteeing thermodynamic consistency and nonnegative entropy generation.

**Corollary (Clausius form).**

If $\sigma = \exp(-\beta H)/Z$, then $\dot{\Sigma}(t) = \dot{S}(\rho_t) - \beta \dot{Q}_t \geq 0$, which recovers the Clausius inequality of the second law of thermodynamics.

**Lemma 5.3.1 (Dual evolution invariance).**

For the adjoint (backward) equation $\dot{E} = -\mathcal{L}^\dagger(E)$, $\frac{d}{dt}\mathrm{Tr}[E(t)\rho(t)] = \mathrm{Tr}[\mathcal{L}(\rho)E] - \mathrm{Tr}[\mathcal{L}^\dagger(E)\rho] = 0$. Thus, the backward informational dynamics preserves the dual pairing and does not affect the physical energy balance. The subsequent propositions and theorems (5.3.1–5.3.8) detail this structure, extending it to multiple reservoirs, continuous monitoring, and feedback. The key ingredients are: (i) well-posedness and complete positivity of the Lindblad generator; (ii) Spohn's inequality (contraction of quantum relative entropy); (iii) Clausius-type bounds for multiple baths; and (iv) the informational (non-energetic) character of the backward equation for $E_t$. Assume a Lindblad generator on $\mathcal{H}$

$$L(\rho) = -\frac{i}{\hbar}[H,\rho] + \sum_j \left(L_j \rho L_j^\dagger - \frac{1}{2}\{L_j^\dagger L_j, \rho\}\right), \tag{5.18}$$

with adjoint

$$L^\dagger(X) = +\frac{i}{\hbar}[H,X] + \sum_j \left(L_j^\dagger X L_j - \frac{1}{2}\{L_j^\dagger L_j, X\}\right). \tag{5.19}$$

Then $e^{tL}$ is CPTP and $e^{tL^\dagger}$ is unital CP (UCP).

**Proposition 5.3.1 (Unitary part leaves von Neumann entropy invariant)**

**Claim.** For the unitary generator $L_H(\rho) = -\frac{i}{\hbar}[H,\rho]$, the von Neumann entropy $S(\rho) = -\mathrm{Tr}[\rho\ln \rho]$ is conserved:

$$\frac{d}{dt}S(\rho_t) = 0.$$

**Proof.** Unitary evolution preserves the spectrum. Directly, $\frac{d}{dt}S(\rho_t) = -\mathrm{Tr}[\dot{\rho}_t \ln \rho_t] = -\frac{i}{\hbar}\mathrm{Tr}\left([H,\rho_t]\ln \rho_t\right) = -\frac{i}{\hbar}\mathrm{Tr}\left(H[\rho_t, \ln \rho_t]\right) = 0$, since $[\rho, \ln \rho] = 0$ and by cyclicity. $\square$

**Theorem 5.3.2 (Spohn's inequality: monotonic decrease of relative entropy)**

**Claim.** Let $\sigma$ be a stationary state of $L$ (i.e., $L(\sigma) = 0$). The quantum relative entropy $D(\rho \parallel \sigma) = \mathrm{Tr}[\rho(\ln \rho - \ln \sigma)]$ is nonincreasing along the Lindblad semigroup:

$$\frac{d}{dt}D(\rho_t \parallel \sigma) = \mathrm{Tr}\left[L(\rho_t)(\ln \rho_t - \ln \sigma)\right] \leq 0. \tag{5.20}$$

**Proof (Spohn).** Differentiate $D$ and use $\partial_t \ln \rho_t = \int_0^\infty (\rho_t + sI)^{-1}\dot{\rho}_t(\rho_t + sI)^{-1}ds$ . The term

$\mathrm{Tr}[\rho_t \, \partial_t \ln \rho_t]$ equals $\mathrm{Tr}[\dot\rho_t] = 0$, hence $\frac{d}{dt} D(\rho_t \parallel \sigma) = \mathrm{Tr}[L(\rho_t)(\ln \rho_t - \ln \sigma)]$. The unitary contribution vanishes by Prop. 5.3.1; the dissipator is $\leq 0$ by operator convexity of $-\ln$ (Klein's inequality), or by differentiating the data-processing inequality $D(\Phi_t(\rho) \parallel \Phi_t(\sigma)) \leq D(\rho \parallel \sigma)$ with $\Phi_t = e^{tL}$.

**Proposition 5.3.3 (Entropy production rate and the second law)**

**Claim.** With $\sigma$ stationary,

$$\Sigma(\rho_t) := -\frac{d}{dt} D(\rho_t \parallel \sigma) = -\mathrm{Tr}\left[L(\rho_t)(\ln \rho_t - \ln \sigma)\right] \geq 0. \tag{5.21}$$

If $\sigma \propto e^{-\beta H}$ (Gibbs), then with heat flow $J_Q(t)$,

$$\dot S(\rho_t) - \beta J_Q(t) = \Sigma(\rho_t) \geq 0, \tag{5.22}$$

i.e., the Clausius inequality.

**Proof.** Nonnegativity follows from (5.3.3). For a Gibbs $\sigma$, write $\ln \sigma = -\beta H - \ln Z$. Using $-\frac{d}{dt} D(\rho_t \parallel \sigma) = -\mathrm{Tr}[\dot\rho_t \ln \rho_t] + \beta \, \mathrm{Tr}[\dot\rho_t H]$, identify $\dot S(\rho_t) = -\mathrm{Tr}[\dot\rho_t \ln \rho_t]$ and set $J_Q(t) := -\mathrm{Tr}[\dot\rho_t H]$ (energy lost by the system = heat to the bath). Then (5.22) follows. For any Lindblad generator $\mathcal L$ with a stationary state $\sigma$ satisfying $\mathcal L(\sigma) = 0$, the quantum relative entropy $D(\rho_t \parallel \sigma) = \mathrm{Tr}[\rho_t(\ln \rho_t - \ln \sigma)]$ is monotonically non-increasing under the dynamics $\dot\rho_t = \mathcal L(\rho_t)$: $\frac{d}{dt} D(\rho_t \parallel \sigma) = \mathrm{Tr}[\mathcal L(\rho_t)(\ln \rho_t - \ln \sigma)] \leq 0$. This ensures that the entropy production rate $\Sigma(\rho_t) := -\frac{d}{dt} D(\rho_t \parallel \sigma) \geq 0$ is always non-negative. In the Gibbs stationary case $\sigma \propto e^{-\beta H}$, this reduces to the Clausius inequality $\dot S - \beta J_Q = \Sigma \geq 0$, demonstrating that the present framework is fully consistent with the second law of thermodynamics.

**Theorem 5.3.4 (Clausius inequality for multiple baths)**

**Claim.** Suppose $L = \sum_r L^{(r)}$ where each $L^{(r)}$ leaves a Gibbs state $\sigma^{(r)} \propto e^{-\beta_r H}$ invariant (Davies limit). Then

$$\dot S(\rho_t) - \sum_r \beta_r J_Q^{(r)}(t) = \sum_r \Sigma^{(r)}(\rho_t) \, \Sigma^{(r)}(\rho_t) \geq 0, \tag{5.23}$$

with $J_Q^{(r)}(t) := -\mathrm{Tr}[L^{(r)}(\rho_t) H]$ and $\Sigma^{(r)}(\rho_t) := -\mathrm{Tr}[L^{(r)}(\rho_t)(\ln \rho_t - \ln \sigma^{(r)})] \geq 0$.

**Proof.** Apply Prop. 5.3.3 to each $r$ and sum. Each term is nonnegative, hence the total is nonnegative. □

**Proposition 5.3.5 (Backward effects do not alter physical thermodynamic balances)**

**Claim.** The backward equation $\dot E_t = -L^\dagger(E_t)$ is informational and does not modify energy/heat/entropy balances. For all $t$,

$$\frac{d}{dt}\mathrm{Tr}[E_t \rho_t] = 0, \tag{5.24}$$

i.e., the dual pairing of $(\rho_t, E_t)$ is conserved; hence introducing $E_t$ leaves the second law intact.

**Proof.** Differentiate: $\frac{d}{dt}\mathrm{Tr}[E_t \rho_t] = \mathrm{Tr}[\dot E_t \rho_t] + \mathrm{Tr}[E_t \dot\rho_t] = \mathrm{Tr}[-L^\dagger(E_t)\rho_t] + \mathrm{Tr}[E_t L(\rho_t)] = 0$, by the adjoint

identity $\text{Tr}[L^\dagger(X)Y] = \text{Tr}[X\,L(Y)]$. □

**Proposition 5.3.6 (Unravellings average back to Lindblad dynamics)**

**Claim.** For diffusive (homodyne) or jump (photon-counting) unravellings (SMEs), the ensemble average reproduces Lindblad dynamics. Hence continuous monitoring is consistent, on average, with the second law (Theorems 5.3.2–5.3.4).

**Proof.** For homodyne (cf. §4(iii)): $d\rho_t = L(\rho_t)\,dt + \sqrt{\eta}\,\mathcal{H}[c](\rho_t)\,dW_t$. Taking expectations, $\mathbb{E}[dW_t] = 0$ gives $\frac{d}{dt}\mathbb{E}[\rho_t] = L(\mathbb{E}[\rho_t])$. Analogous averaging holds for jump trajectories. Thus the mean dynamics satisfy the hypotheses of (5.19)–(5.20). □

**Theorem 5.3.7 (Second law under measurement and feedback: conditional form)**

**Claim.** For conditional states $\rho_t^R$ given a measurement record $R$, one has, in expectation over records,
$$\mathbb{E}_R\left[\dot{S}(\rho_t^R) - \sum_r \beta_r\,J_Q^{(r),R}(t)\right] \geq 0, \tag{5.25}$$
with $J_Q^{(r),R}(t) := -\text{Tr}[L^{(r)}(\rho_t^R)H]$. Introducing conditional backward effects $E_t^R$ does not change the bound.

**Proof.** Linearity of the unravelling implies $\mathbb{E}_R[\rho_t^R] = \rho_t$ and $\mathbb{E}_R[L^{(r)}(\rho_t^R)] = L^{(r)}(\rho_t)$. By convexity (Jensen) and Fatou's lemma, together with Prop. 5.3.6, $\mathbb{E}_R\left[-\frac{d}{dt}D(\rho_t^R \parallel \sigma^{(r)})\right] \geq -\frac{d}{dt}D(\rho_t \parallel \sigma^{(r)}) \geq 0$. Rewriting as in (5.23) yields (5.25). Independence from $E_t^R$ follows from (5.24). □

**Proposition 5.3.8 (Work and the first law: consistent splitting)**

**Claim.** For a time-dependent Hamiltonian $H(t)$ with work rate $\dot{W}(t) = \text{Tr}[\rho_t\,\dot{H}(t)]$, the first law holds:
$$\frac{d}{dt}\langle H\rangle_t = \dot{W}(t) - \sum_r J_Q^{(r)}(t),\quad J_Q^{(r)}(t) := -\text{Tr}\left[H\,L^{(r)}(\rho_t)\right]. \tag{5.26}$$

The pair $(\rho_t, E_t)$ is compatible with this decomposition.

**Proof.** Using $\dot{\rho}_t = L(\rho_t)$, $\frac{d}{dt}\langle H\rangle_t = \text{Tr}[\dot{\rho}_t H] + \text{Tr}[\rho_t \dot{H}] = \sum_r \text{Tr}[L^{(r)}(\rho_t)H] + \dot{W}(t) = \dot{W}(t) - \sum_r J_Q^{(r)}(t)$.

Backward effects do not appear in energy balances (Prop. 5.3.5). □

**Summary of §5.3**

- Relative-entropy contraction (Spohn): Lindblad dynamics yield nonnegative entropy production $\Sigma \geq 0$.
- Single-bath Clausius: $\dot{S} - \beta J_Q = \Sigma \geq 0$; multi-bath generalization holds as a sum.
- Backward effects are informational: the conserved pairing $\frac{d}{dt}\text{Tr}[E_t\rho_t] = 0$ shows $E_t$ does not alter heat or work.
- Continuous monitoring: trajectory averages return to Lindblad; the second law holds on average, and even conditionally in expectation.
- First law compatibility: work and heat splits are preserved.

Consequently, the time-symmetric probabilistic law $\mathrm{Tr}[E\,I(\rho)]$ integrates seamlessly with the first and second laws: introducing future-conditioned backward effects preserves physical bookkeeping of energy and guarantees nonnegative entropy production.

## 5.4 Experimental Consistency

We prove that the bidirectional framework—forward state $\rho_t$, backward effect $E_t$, and instruments $\{I_m\}$—reproduces laboratory statistics across four representative classes: weak measurements, EPR–Bell tests, continuous homodyne monitoring, and photon counting, and that it converges to classical smoothing (Kalman/RTS) in the commutative linear–Gaussian limit.

### 5.4.1 Weak Measurements: First-Order Pointer Shifts with Quadratic Remainder Bounds

**Setting.** System $H_S$ and pointer $H_M$ with $[q,p]=i$ (we set $\hbar=1$). The impulsive coupling $U=\exp(-ig\,A\otimes p)$, $g\ll 1$, is applied to $|i\rangle\otimes|G\rangle$ where $|G\rangle$ is a real Gaussian with $\langle q\rangle_G=\langle p\rangle_G=0$, $\mathrm{Var}(q)=\sigma_q^2$. Post-select on $|f\rangle\langle f|$ and read out the conditional pointer means.

**Theorem 5.4.1 (Weak-value first order with $O(g^2)$ bounds).**

For any bounded $A$ with $\|A\|\le a_{\max}$, the conditional shifts obey

$$\Delta q = g\,\mathrm{Re}\,A_w + O(g^2),\quad \Delta p = 2g\,\mathrm{Var}(p)\,\mathrm{Im}\,A_w + O(g^2),\quad A_w := \frac{\langle f|A|i\rangle}{\langle f|i\rangle}, \tag{5.27}$$

and there exist finite constants $C_q, C_p$ (depending on $a_{\max}$ and the second moments of $|G\rangle$) such that

$$|\Delta q - g\,\mathrm{Re}\,A_w| \le C_q\, g^2,\quad |\Delta p - 2g\,\mathrm{Var}(p)\,\mathrm{Im}\,A_w| \le C_p\, g^2. \tag{5.28}$$

**Proof (sketch).** Expand $U \simeq I - igA\otimes p - \tfrac{1}{2}g^2 A^2 \otimes p^2 + \cdots$.

The unnormalized pointer state is

$$\langle f|U|i\rangle|G\rangle \simeq \langle f|i\rangle(I - igA_w p - \tfrac{1}{2}g^2 B_w p^2 + \cdots)|G\rangle,\quad B_w := \frac{\langle f|A^2|i\rangle}{\langle f|i\rangle}\ldots \tag{5.29}$$

The normalization is $|\langle f|i\rangle|^2[1 + g^2\mathrm{Var}(p)(|A_w|^2 - \mathrm{Re}\,B_w) + O(g^3)]$. Using $\langle[q,p]\rangle = i$ and $\langle\{q,p\}\rangle = 0$, one obtains (5.27); (5.28) follows by bounding the $O(g^2)$ terms with $\|A\|, \langle p^2\rangle_G$.

**Experimental implication.** Optical weak measurements (balanced homodyne with small phase), Rydberg probes, and CV platforms report pointer means consistent with (5.4.1). Multiple post-selections and composite observables follow by composing Kraus instruments.

### 5.4.2 EPR–Bell: Maximal CHSH Violation with Uniform Nonselective Marginals

**Setting.** Bell state $|\Phi^+\rangle = (|00\rangle + |11\rangle)/\sqrt{2}$. Alice and Bob perform $\pm 1$ observables

$$P_{\mathbf{a}}^\pm = \frac{I \pm \mathbf{a}\cdot\boldsymbol{\sigma}}{2},\quad P_{\mathbf{b}}^\pm = \frac{I \pm \mathbf{b}\cdot\boldsymbol{\sigma}}{2}.$$

**Theorem 5.4.2 (CHSH maximum and uniform marginals).**

(i) The joint distribution

$$p(s,t) = \mathrm{Tr}[(P_{\mathbf{a}}^s \otimes P_{\mathbf{b}}^t)|\Phi^+\rangle\langle\Phi^+|] = \frac{1 + st\,\mathbf{a}\cdot\mathbf{b}}{4} \tag{5.30}$$

yields the correlation $E(\mathbf{a}, \mathbf{b}) = \mathbf{a} \cdot \mathbf{b}$ and the Tsirelson bound $S = 2\sqrt{2}$ for appropriate settings.

(ii) The nonselective marginal is uniform:

$$p_B(\pm) = \sum_{s=\pm} p(s, \pm) = \tfrac{1}{2}, \tag{5.31}$$

independent of Alice's basis, operation, or selection.

**Proof.** Use $\langle \sigma_i \otimes \sigma_j \rangle_{\Phi^+} = \delta_{ij}$ and $\sum_s P_{\mathbf{a}}^s = I$.

**Experimental implication.** Photon polarization, trapped ions, and superconducting qubits realize (5.30)–(5.31). Selective post-analysis on $A$ leaves the $B$-marginal unchanged when nonselective, in agreement with §5.2.

### 5.4.3 Continuous Homodyne: Path Likelihood, Past Quantum State, and Martingale Innovations

**Setting.** Homodyne detection with rate $\kappa$, efficiency $\eta$, observable $X_c = c + c^\dagger$. The SME and current readout are

$$d\rho_t = \mathcal{L}(\rho_t)\, dt + \sqrt{\eta\kappa}\, \mathcal{H}[c](\rho_t)\, dW_t,\, dY_t = 2\sqrt{\eta\kappa}\, \langle X_c \rangle_t\, dt + dW_t. \tag{5.32}$$

**Theorem 5.4.3 (Trajectory likelihood and PQS equivalence).**

Up to normalization, the likelihood density of the record $\{Y_s\}_{0 \leq s \leq t_f}$ is

$$\mathcal{L}[\{Y\}] \propto \exp\left[-\tfrac{1}{2} \int_0^{t_f}(dY_t - 2\sqrt{\eta\kappa}\, \langle X_c \rangle_t\, dt)^2\right]. \tag{5.33}$$

With the adjoint SDE

$$dE_t = -\mathcal{L}^\dagger(E_t)\, dt + \sqrt{\eta\kappa}\, \mathcal{H}^\dagger[c](E_t)\, dW_t,\, E_{t_f} = E_f, \tag{5.34}$$

the conditional probability of any intermediate POVM $\{M_m\}$ is

$$p(m \mid \{Y\}_{0:t_f}) = \frac{\text{Tr}\,[E(t^+)\, I_m\, (\rho(t^-))]}{\sum_k \text{Tr}\,[E(t^+)\, I_k\, (\rho(t^-))]}. \tag{5.35}$$

**Proof (sketch).** Girsanov's theorem yields (5.34). Continuous-time Bayes/Radon–Nikodym updates generate the $\mathcal{H}[c]$ innovation term forward and $\mathcal{H}^\dagger[c]$ backward, giving (5.35), which matches the unified law.

**Proposition 5.4.4 (Innovation martingale).**

$$d\widetilde{W}_t := dY_t - 2\sqrt{\eta\kappa}\, \langle X_c \rangle_t\, dt \tag{5.36}$$

is a martingale with respect to the measurement filtration: $\mathbb{E}[d\widetilde{W}_t \mid \mathcal{F}_t] = 0$.

**Experimental implication.** Quantum-trajectory tomography in superconducting circuits and cavity QED implements smoothing exactly as (5.35); intermediate retrodictive probabilities follow in one line.

### 5.4.4 Photon Counting: Jump Trajectories and Retrodictive Probabilities

**Setting.** Ideal photon counting for jump operator $c$ (no dark counts). Discrete time step $\Delta t$, event $n_t \in \{0,1\}$.

**Theorem 5.4.5 (Jump SME and past quantum state).**

Forward conditional update:

$$\rho_{t+\Delta t} \propto \begin{cases} c\, \rho_t\, c^\dagger\, \Delta t, & n_t = 1, \\ (I - \tfrac{1}{2} c^\dagger c\, \Delta t)\rho_t(I - \tfrac{1}{2} c^\dagger c\, \Delta t), & n_t = 0, \end{cases} \tag{5.37}$$

Backward adjoint update:

$$E_t \propto \begin{cases} c^\dagger E_{t+\Delta t}\, c, & n_t = 1, \\ (I - \tfrac{1}{2} c^\dagger c\, \Delta t)E_{t+\Delta t}(I - \tfrac{1}{2} c^\dagger c\, \Delta t), & n_t = 0, \end{cases} \tag{5.38}$$

and the probability of any inserted intermediate POVM is given by the unified formula (5.35).

**Proof (sketch).** The Poisson likelihood $(\lambda_t \Delta t)^{n_t} e^{-\lambda_t \Delta t}$ with $\lambda_t = \text{Tr}[c^\dagger c \rho_t]$, combined with Bayes updates, yields (5.37)–(5.38). Factorization over time and final reduction produce (5.35). □

**Experimental implication.** Single-photon detection and quantum-jump experiments match these statistics; inefficiencies/backgrounds enter as instrument parameters.

### 5.4.5 Classical Limit: Kalman/RTS Smoothing and the Triple Product

**Theorem 5.4.6 (Linear–Gaussian correspondence).**

For the commutative linear–Gaussian model

$$\dot{x}_t = A x_t + \xi_t, \quad y_t = C x_t + v_t, \quad \xi \sim \mathcal{N}(0, Q), \quad v \sim \mathcal{N}(0, R), \qquad \text{......(5.39)}$$

the smoothed probability of a discrete event $m$ at time $t$ has the triple-product form

$$p(m|\text{all data}) \propto \underbrace{\beta_{t^+}(x)}_{\text{RTS backward}} \; \underbrace{\mathcal{N}(m; Hx, \Sigma)}_{\text{likelyhood}} \; \underbrace{\pi_{t^-}(x)}_{\text{Kalman forward}}, \qquad \text{......(5.40)}$$

which is exactly the classical limit of the quantum law $\text{Tr}[E(t^+) I_m(\rho(t^-))]$ under the replacements $\rho \to \pi, E \to \beta, I_m \to L_m$.

**Proof.** In the commutative subalgebra, traces become integrals/sums; the instrument reduces to a likelihood $L_m$. Linear–Gaussian closure gives Kalman forward $\pi$ and RTS backward $\beta$, yielding (5.40). □

**Experimental implication.** In the large-photon-number/large-LO limit, circuit QED and optical records reduce to Kalman smoothing, confirming continuity with classical estimation.

**Summary of §5.4**

- Weak measurements: First-order pointer laws with explicit $O(g^2)$ control (Thm. 5.4.1).
- EPR–Bell: Maximal CHSH violation with uniform nonselective marginals (Thm. 5.4.2).
- Continuous homodyne: Path-likelihood and PQS coincide; innovation is a martingale (Thm. 5.4.3, Prop. 5.4.4).
- Photon counting: Jump SMEs and adjoint updates recover unified retrodictive probabilities (Thm. 5.4.5).
- Classical limit: The quantum rule reduces to the Kalman/RTS triple product (Thm. 5.4.6).

Thus, across discrete/continuous, weak/strong, and Gaussian/non-Gaussian regimes, experimental statistics are uniformly captured by the single time-symmetric law

$$p(m \mid i, f) = \frac{\text{Tr}[E \, I_m(\rho)]}{\sum_k \text{Tr}[E \, I_k(\rho)]},$$

establishing comprehensive empirical consistency.

## 6 Discussion

This work redefines quantum measurement not as a fundamentally time-asymmetric process, but as a time-symmetric rule for updating information. The reformulation reorganizes the relation among *time, causality,* and *information* within quantum theory: without abandoning any physical content, it reconstructs time symmetry from the intrinsic duality of operator algebras. The result is a framework in which dynamical laws, probabilities, and thermodynamic constraints are co-expressed through a single, self-consistent operator calculus.

**6.1 Theoretical significance: redefining the arrow of time**

Although the basic dynamical equations of quantum theory (Schrödinger/von Neumann) are time-reversal symmetric, the measurement update has long appeared intrinsically one-way. Traditional accounts—state "collapse," dynamical decoherence—retain an implicit directional bias in information flow. Here, the pair of symmetric evolution equations, $\dot{\rho}_t = L(\rho_t)$, $\dot{E}_t = -L^\dagger(E_t)$, shifts the asymmetry from *physical law* to the *observer's information state*. Within this view, causality, thermodynamics, and observational statistics sit on the same operator-theoretic footing, and the "arrow of time" emerges as an arrow of information. Future data may consistently condition present inferences without invoking retrocausal dynamics: the process is better understood as temporal information fusion, not as signals traveling backward in time.

**6.2 Relation to existing time-symmetric and informational approaches**

Our account is aligned with, yet decisively distinct from, TSVF, the transactional interpretation, and QBism.

- TSVF. Rather than personifying the bra as a "wave from the future," we treat $E_t$ as an *effect operator* propagated by the adjoint channel. This eliminates interpretational tensions—"retro-particles," causal reversals—while preserving the operational content of pre- and post-selection.
- Transactional interpretation. What TI frames as "offer" and "confirmation" waves is recast as a dual pairing of probabilistic weights. Because no physical signal is required to propagate from the future, conflicts with relativistic causality are removed.
- QBism. While sharing an emphasis on probabilistic updating, our formulation is fully operational and observer-independent: updates are fixed by the adjointness of CPTP maps, not by subjective degrees of belief. In this sense, the present framework unifies time-symmetric intuitions with strict mathematical and physical consistency.

**6.3 Thermodynamic consistency and entropy production**

As established in §5, Spohn's inequality for the Lindblad semigroup, $\frac{d}{dt}D(\rho_t \| \sigma) \leq 0$, connects continuously to the Boltzmann–Kullback inequality in the classical limit. Entropy production thus acquires a purely informational meaning: it quantifies the irretrievable loss or dispersion of information through open-system dynamics and measurement coarsening. The second law and quantum measurement theory become two facets of the same information-theoretic structure. The "arrow of time" is then explained as the irreversibility of information flow, not as a fundamental asymmetry of microscopic dynamics.

**6.4 Classical limit and probabilistic coherence**

In the limit $\hbar \to 0$, the time-symmetric formalism reproduces Bayesian smoothing (RTS) and the backward–forward structure of the Kalman filter. Quantum and classical inference are thereby recognized as sharing a common architecture: a fusion of forward prediction with backward information refinement. This continuity rigorously supports the view that quantum probability extends classical probability without rupture, enabling a seamless bridge among quantum measurement, classical estimation, and stochastic thermodynamics. The framework thus qualifies as a unified foundational theory across these regimes.

**6.5 Outlook**

Several directions naturally extend the present results.

1. Non-Markovian generalizations. Incorporate memory kernels and environment correlations so that $E_t$ acquires explicit history dependence while preserving complete positivity and adjoint consistency.
2. Quantum control and feedback. Exploit bidirectional inference for *closed-loop* design—weak measurements, homodyne monitoring, and real-time control—to optimize performance under realistic noise and constraints.
3. Quantum thermomachinery. Track energy and entropy as *information flows* within engines and refrigerators, formulating fully time-symmetric, thermodynamically consistent bounds for performance and irreversibility.
4. Information-theoretic quantum gravity. Explore how the bidirectional informational structure interfaces with entanglement geometry (e.g., AdS/CFT) and ER=EPR-like correspondences, where dual descriptions and boundary conditions play a central role.

These avenues promise new contact points among quantum information, statistical physics, and fundamental theory.

**6.6 Conclusion: time as information**

Ultimately, the coexistence of *reversible* microscopic laws with *irreversible* observations is not a paradox of time itself, but a property of asymmetric information allocation. The arrow of time simply records the direction in which information is *conditioned*. By reframing quantum theory as a probabilistic information theory—in which observation, inference, and thermodynamics share a common operator-dual backbone—we expose a latent "information spacetime geometry" within the mathematics of quantum mechanics. This perspective delivers a coherent, testable, and thermodynamically lawful account of time-symmetric measurement, and it grounds a program for unifying inference across quantum and classical worlds without sacrificing physical realism.

**7 Conclusion**

This study has re-examined the intrinsic time asymmetry of quantum measurement and redefined it not as *physical irreversibility* but as an *informational asymmetry*. Whereas conventional interpretations rely on wavefunction collapse or decoherence to explain the unidirectionality of measurement, the present theory reconstructs the process as a time-symmetric information update governed by a pair of dual evolution equations based on a completely positive map $L$ and its adjoint $L^\dagger$: $\dot\rho_t = L(\rho_t), \dot E_t = -L^\dagger(E_t)$.

The forward state $\rho_t$ and backward effect $E_t$ thus represent complementary streams of information, whose intersection term $\mathrm{Tr}[E\, I_m(\rho)]$ defines the conditional probability $p(m \mid i, f)$ without presupposing any temporal direction.

Without invoking time-reversal operators or hypothetical "retrocausal particles," this dual structure preserves causality, complete positivity, and no-signaling while extending the statistical architecture of quantum measurement into a time-symmetric form. The adjoint relation between $L$ and $L^\dagger$ ensures full thermodynamic consistency: through Spohn's inequality, the framework aligns with the law of nonnegative entropy production and, in the classical limit, converges continuously to the Fokker–Planck equation and Bayesian smoothing probabilities.

Consequently, quantum, classical, and thermodynamic descriptions are unified within a single operator-theoretic framework representing bidirectional information flow. The arrow of time thereby emerges not as a property of physical law but as a manifestation of the observer's informational asymmetry. Observation itself is recast as an operational process that fuses past and future information to condition the present state, producing a statistically closed description of reality.

The significance of this framework lies in its ability to liberate quantum measurement from causal irreversibility and to link time, information, and entropy through a shared mathematical foundation. The longstanding dichotomy between reversible physical laws and irreversible observation is resolved by interpreting temporal asymmetry as a *bias in informational flow*. Quantum theory thus ceases to be merely a theory of matter and becomes a theory of bidirectional information dynamics.

Ultimately, this perspective points toward a new unifying principle across quantum information, thermodynamics, and statistical physics—one in which "Time ≡ Asymmetry of Information." This insight establishes a foundation for a new physical paradigm: a world governed not by the one-way evolution of matter, but by the symmetric exchange and conditioning of information.

**References**

1. Horowitz, J. M., & Esposito, M. (2014). *Thermodynamics with continuous information flow.* Physical Review X, 4(3), 031015.
2. Dewar, R. (2003). *Information theory explanation of the fluctuation theorem, maximum entropy production and self-organized criticality in non-equilibrium stationary states.* Journal of Physics A: Mathematical and General, 36(3), 631–641.
3. Niven, R. K. (2010). *Steady-state generalization of the Boltzmann distribution.* Physical Review E, 82(1), 011109.
4. England, J. L. (2020). *Statistical physics of self-replication.* The Journal of Chemical Physics, 152(13), 130901.
5. Aharonov, Y., Bergmann, P. G., & Lebowitz, J. L. (1964). *Time symmetry in the quantum process of measurement.* Physical Review, 134(6B), B1410–B1416.
6. Aharonov, Y., & Vaidman, L. (1990). *Properties of a quantum system during the time interval between two measurements.* Physical Review A, 41(1), 11–20.
7. Cramer, J. G. (1986). *The transactional interpretation of quantum mechanics.* Reviews of Modern Physics, 58(3), 647–688.
8. Fuchs, C. A., & Schack, R. (2013). *Quantum-Bayesian coherence.* Reviews of Modern Physics, 85(4), 1693–1715.
9. Davies, E. B., & Ozawa, M. (1979). *Quantum dynamical semigroups and the operation-valued measure.* Communications in Mathematical Physics, 63(3), 281–305.
10. Lindblad, G. (1976). *On the generators of quantum dynamical semigroups.* Communications in Mathematical Physics, 48(2), 119–130.

11. Spohn, H. (1978). *Entropy production for quantum dynamical semigroups.* Journal of Mathematical Physics, 19(5), 1227–1230.
12. Klein, O. (1931). *Zur Quantenmechanischen Begründung des zweiten Hauptsatzes der Wärmetheorie.* Zeitschrift für Physik, 72(11–12), 767–775.
13. Clausius, R. (1865). *Über verschiedene für die Anwendung bequeme Formen der Hauptgleichungen der mechanischen Wärmetheorie.* Annalen der Physik, 201(7), 353–400.
14. Davies, E. B. (1974). *Markovian master equations.* Communications in Mathematical Physics, 39(2), 91–110.
15. Tomonaga, S. (1946). *On a relativistically invariant formulation of the quantum theory of wave fields.* Progress of Theoretical Physics, 1(2), 27–42.
16. Schwinger, J. (1948). *Quantum electrodynamics. I. A covariant formulation.* Physical Review, 74(10), 1439–1461.
17. Wiseman, H. M. (1996). *Quantum trajectories and feedback.* Quantum and Semiclassical Optics: Journal of the European Optical Society Part B, 8(1), 205–222.
18. Jacobs, K., & Steck, D. A. (2006). *A straightforward introduction to continuous quantum measurement.* Contemporary Physics, 47(5), 279–303.
19. Breuer, H.-P., & Petruccione, F. (2002). *The Theory of Open Quantum Systems.* Oxford University Press, Oxford.
20. von Neumann, J. (1955). *Mathematical Foundations of Quantum Mechanics.* Princeton University Press, Princeton.
21. Born, M. (1926). *Zur Quantenmechanik der Stoßvorgänge.* Zeitschrift für Physik, 37(12), 863–867.
22. Bayes, T., & Price, R. (1763). *An essay towards solving a problem in the doctrine of chances.* Philosophical Transactions of the Royal Society of London, 53, 370–418.
23. Kolmogorov, A. N. (1933). *Foundations of the Theory of Probability.* Chelsea Publishing Company, New York.
24. Kalman, R. E. (1960). *A new approach to linear filtering and prediction problems.* Journal of Basic Engineering, 82(1), 35–45.
25. Rauch, H. E., Tung, F., & Striebel, C. T. (1965). *Maximum likelihood estimates of linear dynamic systems.* AIAA Journal, 3(8), 1445–1450.
26. Prigogine, I. (1978). *Time, structure and fluctuations.* Science, 201(4358), 777–785.
27. Jaynes, E. T. (1957). *Information theory and statistical mechanics.* Physical Review, 106(4), 620–630.
28. Kullback, S., & Leibler, R. A. (1951). *On information and sufficiency.* Annals of Mathematical Statistics, 22(1), 79–86.